\documentclass[twocolumn,superscriptaddress,prd,showkeys,showpacs,nofootinbib]{revtex4-1}

\usepackage{graphicx}
\usepackage[colorlinks = true,
            linkcolor = cyan,
            urlcolor  = blue,
            citecolor = red,
            anchorcolor = blue]{hyperref}\usepackage{color}
\usepackage{amssymb}
\usepackage{amsthm}
\usepackage{textcomp}
\usepackage{mathtools}
\usepackage{comment}
\usepackage[separate-uncertainty,retain-explicit-plus,per-mode = symbol]{siunitx}
\usepackage{url}

\newcommand{\dl}{\mbox{DAMA/LIBRA}}
\newcommand{\dn}{\mbox{DAMA/NaI}}
\newcommand{\cogent}{\mbox{CoGeNT}}
\newcommand{\nai}{\mbox{NaI(Tl)}}
\newcommand{\hthree}{\mbox{$^{3}$H}}
\newcommand{\cfour}{\mbox{$^{14}$C}}

\newcommand{\krthree}{\mbox{$^{83m}$Kr}}

\newcommand{\cftwo}{\mbox{$^{252}$Cf}}
\newcommand{\arsev}{\mbox{$^{37}$Ar}}

\newcommand{\xesev}{\mbox{$^{127}$Xe}}

\newcommand{\keVee}{\mbox{keV$_{\text{ee}}$}}

\begin{document}

%\preprint{APS/123-QED}
\title{Search for annual and diurnal rate modulations in the LUX experiment}

\author{D.S.~Akerib} \affiliation{Case Western Reserve University, Department of Physics, 10900 Euclid Ave, Cleveland, OH 44106, USA} \affiliation{SLAC National Accelerator Laboratory, 2575 Sand Hill Road, Menlo Park, CA 94205, USA} \affiliation{Kavli Institute for Particle Astrophysics and Cosmology, Stanford University, 452 Lomita Mall, Stanford, CA 94309, USA}
\author{S.~Alsum} \affiliation{University of Wisconsin-Madison, Department of Physics, 1150 University Ave., Madison, WI 53706, USA}  
\author{H.M.~Ara\'{u}jo} \affiliation{Imperial College London, High Energy Physics, Blackett Laboratory, London SW7 2BZ, United Kingdom}  
\author{X.~Bai} \affiliation{South Dakota School of Mines and Technology, 501 East St Joseph St., Rapid City, SD 57701, USA}  
\author{J.~Balajthy} \affiliation{University of California Davis, Department of Physics, One Shields Ave., Davis, CA 95616, USA}  
\author{P.~Beltrame} \affiliation{SUPA, School of Physics and Astronomy, University of Edinburgh, Edinburgh EH9 3FD, United Kingdom}  
\author{E.P.~Bernard} \affiliation{University of California Berkeley, Department of Physics, Berkeley, CA 94720, USA}  
\author{A.~Bernstein} \affiliation{Lawrence Livermore National Laboratory, 7000 East Ave., Livermore, CA 94551, USA}  
\author{T.P.~Biesiadzinski} \affiliation{Case Western Reserve University, Department of Physics, 10900 Euclid Ave, Cleveland, OH 44106, USA} \affiliation{SLAC National Accelerator Laboratory, 2575 Sand Hill Road, Menlo Park, CA 94205, USA} \affiliation{Kavli Institute for Particle Astrophysics and Cosmology, Stanford University, 452 Lomita Mall, Stanford, CA 94309, USA}
\author{E.M.~Boulton} \affiliation{University of California Berkeley, Department of Physics, Berkeley, CA 94720, USA} \affiliation{Lawrence Berkeley National Laboratory, 1 Cyclotron Rd., Berkeley, CA 94720, USA} \affiliation{Yale University, Department of Physics, 217 Prospect St., New Haven, CT 06511, USA}
\author{B.~Boxer} \affiliation{University of Liverpool, Department of Physics, Liverpool L69 7ZE, UK}  
\author{P.~Br\'as} \affiliation{LIP-Coimbra, Department of Physics, University of Coimbra, Rua Larga, 3004-516 Coimbra, Portugal}  
\author{S.~Burdin} \affiliation{University of Liverpool, Department of Physics, Liverpool L69 7ZE, UK}  
\author{D.~Byram} \affiliation{University of South Dakota, Department of Physics, 414E Clark St., Vermillion, SD 57069, USA} \affiliation{South Dakota Science and Technology Authority, Sanford Underground Research Facility, Lead, SD 57754, USA} 
\author{M.C.~Carmona-Benitez} \affiliation{Pennsylvania State University, Department of Physics, 104 Davey Lab, University Park, PA  16802-6300, USA}  
\author{C.~Chan} \affiliation{Brown University, Department of Physics, 182 Hope St., Providence, RI 02912, USA}  
\author{J.E.~Cutter} \affiliation{University of California Davis, Department of Physics, One Shields Ave., Davis, CA 95616, USA}  
\author{T.J.R.~Davison} \affiliation{SUPA, School of Physics and Astronomy, University of Edinburgh, Edinburgh EH9 3FD, United Kingdom}  
\author{E.~Druszkiewicz} \affiliation{University of Rochester, Department of Physics and Astronomy, Rochester, NY 14627, USA}  
\author{S.R.~Fallon} \affiliation{University at Albany, State University of New York, Department of Physics, 1400 Washington Ave., Albany, NY 12222, USA}  
\author{A.~Fan} \affiliation{SLAC National Accelerator Laboratory, 2575 Sand Hill Road, Menlo Park, CA 94205, USA} \affiliation{Kavli Institute for Particle Astrophysics and Cosmology, Stanford University, 452 Lomita Mall, Stanford, CA 94309, USA} 
\author{S.~Fiorucci} \affiliation{Lawrence Berkeley National Laboratory, 1 Cyclotron Rd., Berkeley, CA 94720, USA} \affiliation{Brown University, Department of Physics, 182 Hope St., Providence, RI 02912, USA} 
\author{R.J.~Gaitskell} \affiliation{Brown University, Department of Physics, 182 Hope St., Providence, RI 02912, USA}  
\author{J.~Genovesi} \affiliation{University at Albany, State University of New York, Department of Physics, 1400 Washington Ave., Albany, NY 12222, USA}  
\author{C.~Ghag} \affiliation{Department of Physics and Astronomy, University College London, Gower Street, London WC1E 6BT, United Kingdom}  
\author{M.G.D.~Gilchriese} \affiliation{Lawrence Berkeley National Laboratory, 1 Cyclotron Rd., Berkeley, CA 94720, USA}  
\author{C.~Gwilliam} \affiliation{University of Liverpool, Department of Physics, Liverpool L69 7ZE, UK}  
\author{C.R.~Hall} \affiliation{University of Maryland, Department of Physics, College Park, MD 20742, USA}  
\author{S.J.~Haselschwardt} \affiliation{University of California Santa Barbara, Department of Physics, Santa Barbara, CA 93106, USA}  
\author{S.A.~Hertel} \affiliation{University of Massachusetts, Amherst Center for Fundamental Interactions and Department of Physics, Amherst, MA 01003-9337 USA} \affiliation{Lawrence Berkeley National Laboratory, 1 Cyclotron Rd., Berkeley, CA 94720, USA} 
\author{D.P.~Hogan} \affiliation{University of California Berkeley, Department of Physics, Berkeley, CA 94720, USA}  
\author{M.~Horn} \affiliation{South Dakota Science and Technology Authority, Sanford Underground Research Facility, Lead, SD 57754, USA} \affiliation{University of California Berkeley, Department of Physics, Berkeley, CA 94720, USA} 
\author{D.Q.~Huang} \affiliation{Brown University, Department of Physics, 182 Hope St., Providence, RI 02912, USA}  
\author{C.M.~Ignarra} \affiliation{SLAC National Accelerator Laboratory, 2575 Sand Hill Road, Menlo Park, CA 94205, USA} \affiliation{Kavli Institute for Particle Astrophysics and Cosmology, Stanford University, 452 Lomita Mall, Stanford, CA 94309, USA} 
\author{R.G.~Jacobsen} \affiliation{University of California Berkeley, Department of Physics, Berkeley, CA 94720, USA}  
\author{W.~Ji} \affiliation{Case Western Reserve University, Department of Physics, 10900 Euclid Ave, Cleveland, OH 44106, USA} \affiliation{SLAC National Accelerator Laboratory, 2575 Sand Hill Road, Menlo Park, CA 94205, USA} \affiliation{Kavli Institute for Particle Astrophysics and Cosmology, Stanford University, 452 Lomita Mall, Stanford, CA 94309, USA}
\author{K.~Kamdin} \affiliation{University of California Berkeley, Department of Physics, Berkeley, CA 94720, USA} \affiliation{Lawrence Berkeley National Laboratory, 1 Cyclotron Rd., Berkeley, CA 94720, USA} 
\author{K.~Kazkaz} \affiliation{Lawrence Livermore National Laboratory, 7000 East Ave., Livermore, CA 94551, USA}  
\author{D.~Khaitan} \affiliation{University of Rochester, Department of Physics and Astronomy, Rochester, NY 14627, USA}  
\author{R.~Knoche} \affiliation{University of Maryland, Department of Physics, College Park, MD 20742, USA}  
\author{E.V.~Korolkova} \affiliation{University of Sheffield, Department of Physics and Astronomy, Sheffield, S3 7RH, United Kingdom}  
\author{S.~Kravitz} \affiliation{Lawrence Berkeley National Laboratory, 1 Cyclotron Rd., Berkeley, CA 94720, USA}  
\author{V.A.~Kudryavtsev} \affiliation{University of Sheffield, Department of Physics and Astronomy, Sheffield, S3 7RH, United Kingdom}  
\author{B.G.~Lenardo} \affiliation{University of California Davis, Department of Physics, One Shields Ave., Davis, CA 95616, USA} \affiliation{Lawrence Livermore National Laboratory, 7000 East Ave., Livermore, CA 94551, USA} 
\author{K.T.~Lesko} \affiliation{Lawrence Berkeley National Laboratory, 1 Cyclotron Rd., Berkeley, CA 94720, USA}  
\author{J.~Liao} \affiliation{Brown University, Department of Physics, 182 Hope St., Providence, RI 02912, USA}  
\author{J.~Lin} \affiliation{University of California Berkeley, Department of Physics, Berkeley, CA 94720, USA}  \affiliation{Lawrence Berkeley National Laboratory, 1 Cyclotron Rd., Berkeley, CA 94720, USA}
\author{A.~Lindote} \affiliation{LIP-Coimbra, Department of Physics, University of Coimbra, Rua Larga, 3004-516 Coimbra, Portugal}  
\author{M.I.~Lopes} \affiliation{LIP-Coimbra, Department of Physics, University of Coimbra, Rua Larga, 3004-516 Coimbra, Portugal}  
\author{A.~Manalaysay} \affiliation{University of California Davis, Department of Physics, One Shields Ave., Davis, CA 95616, USA}  
\author{R.L.~Mannino} \affiliation{Texas A \& M University, Department of Physics, College Station, TX 77843, USA} \affiliation{University of Wisconsin-Madison, Department of Physics, 1150 University Ave., Madison, WI 53706, USA} 
\author{N.~Marangou} \affiliation{Imperial College London, High Energy Physics, Blackett Laboratory, London SW7 2BZ, United Kingdom}  
\author{M.F.~Marzioni} \affiliation{SUPA, School of Physics and Astronomy, University of Edinburgh, Edinburgh EH9 3FD, United Kingdom}  
\author{D.N.~McKinsey} \affiliation{University of California Berkeley, Department of Physics, Berkeley, CA 94720, USA} \affiliation{Lawrence Berkeley National Laboratory, 1 Cyclotron Rd., Berkeley, CA 94720, USA} 
\author{D.-M.~Mei} \affiliation{University of South Dakota, Department of Physics, 414E Clark St., Vermillion, SD 57069, USA}  
\author{M.~Moongweluwan} \affiliation{University of Rochester, Department of Physics and Astronomy, Rochester, NY 14627, USA}  
\author{J.A.~Morad} \affiliation{University of California Davis, Department of Physics, One Shields Ave., Davis, CA 95616, USA}  
\author{A.St.J.~Murphy} \affiliation{SUPA, School of Physics and Astronomy, University of Edinburgh, Edinburgh EH9 3FD, United Kingdom}  
\author{C.~Nehrkorn} \affiliation{University of California Santa Barbara, Department of Physics, Santa Barbara, CA 93106, USA}  
\author{H.N.~Nelson} \affiliation{University of California Santa Barbara, Department of Physics, Santa Barbara, CA 93106, USA}  
\author{F.~Neves} \affiliation{LIP-Coimbra, Department of Physics, University of Coimbra, Rua Larga, 3004-516 Coimbra, Portugal}  
\author{K.C.~Oliver-Mallory} \affiliation{University of California Berkeley, Department of Physics, Berkeley, CA 94720, USA} \affiliation{Lawrence Berkeley National Laboratory, 1 Cyclotron Rd., Berkeley, CA 94720, USA} 
\author{K.J.~Palladino} \affiliation{University of Wisconsin-Madison, Department of Physics, 1150 University Ave., Madison, WI 53706, USA}  
\author{E.K.~Pease} \affiliation{University of California Berkeley, Department of Physics, Berkeley, CA 94720, USA} \affiliation{Lawrence Berkeley National Laboratory, 1 Cyclotron Rd., Berkeley, CA 94720, USA} 
\author{G.R.C.~Rischbieter} \affiliation{University at Albany, State University of New York, Department of Physics, 1400 Washington Ave., Albany, NY 12222, USA}  
\author{C.~Rhyne} \affiliation{Brown University, Department of Physics, 182 Hope St., Providence, RI 02912, USA}  
\author{P.~Rossiter} \affiliation{University of Sheffield, Department of Physics and Astronomy, Sheffield, S3 7RH, United Kingdom}  
\author{S.~Shaw} \affiliation{University of California Santa Barbara, Department of Physics, Santa Barbara, CA 93106, USA} \affiliation{Department of Physics and Astronomy, University College London, Gower Street, London WC1E 6BT, United Kingdom} 
\author{T.A.~Shutt} \affiliation{Case Western Reserve University, Department of Physics, 10900 Euclid Ave, Cleveland, OH 44106, USA} \affiliation{SLAC National Accelerator Laboratory, 2575 Sand Hill Road, Menlo Park, CA 94205, USA} \affiliation{Kavli Institute for Particle Astrophysics and Cosmology, Stanford University, 452 Lomita Mall, Stanford, CA 94309, USA}
\author{C.~Silva} \affiliation{LIP-Coimbra, Department of Physics, University of Coimbra, Rua Larga, 3004-516 Coimbra, Portugal}  
\author{M.~Solmaz} \affiliation{University of California Santa Barbara, Department of Physics, Santa Barbara, CA 93106, USA}  
\author{V.N.~Solovov} \affiliation{LIP-Coimbra, Department of Physics, University of Coimbra, Rua Larga, 3004-516 Coimbra, Portugal}  
\author{P.~Sorensen} \affiliation{Lawrence Berkeley National Laboratory, 1 Cyclotron Rd., Berkeley, CA 94720, USA}  
\author{T.J.~Sumner} \affiliation{Imperial College London, High Energy Physics, Blackett Laboratory, London SW7 2BZ, United Kingdom}  
\author{M.~Szydagis} \affiliation{University at Albany, State University of New York, Department of Physics, 1400 Washington Ave., Albany, NY 12222, USA}  
\author{D.J.~Taylor} \affiliation{South Dakota Science and Technology Authority, Sanford Underground Research Facility, Lead, SD 57754, USA}  
\author{W.C.~Taylor} \affiliation{Brown University, Department of Physics, 182 Hope St., Providence, RI 02912, USA}  
\author{B.P.~Tennyson} \affiliation{Yale University, Department of Physics, 217 Prospect St., New Haven, CT 06511, USA}  
\author{P.A.~Terman} \affiliation{Texas A \& M University, Department of Physics, College Station, TX 77843, USA}  
\author{D.R.~Tiedt} \affiliation{South Dakota School of Mines and Technology, 501 East St Joseph St., Rapid City, SD 57701, USA}  
\author{W.H.~To} \affiliation{California State University Stanislaus, Department of Physics, 1 University Circle, Turlock, CA 95382, USA}  
\author{M.~Tripathi} \affiliation{University of California Davis, Department of Physics, One Shields Ave., Davis, CA 95616, USA}  
\author{L.~Tvrznikova} \affiliation{University of California Berkeley, Department of Physics, Berkeley, CA 94720, USA} \affiliation{Lawrence Berkeley National Laboratory, 1 Cyclotron Rd., Berkeley, CA 94720, USA} \affiliation{Yale University, Department of Physics, 217 Prospect St., New Haven, CT 06511, USA}
\author{U.~Utku} \affiliation{Department of Physics and Astronomy, University College London, Gower Street, London WC1E 6BT, United Kingdom}  
\author{S.~Uvarov} \affiliation{University of California Davis, Department of Physics, One Shields Ave., Davis, CA 95616, USA}  
\author{V.~Velan} \affiliation{University of California Berkeley, Department of Physics, Berkeley, CA 94720, USA}  
\author{J.R.~Verbus} \affiliation{Brown University, Department of Physics, 182 Hope St., Providence, RI 02912, USA}  
\author{R.C.~Webb} \affiliation{Texas A \& M University, Department of Physics, College Station, TX 77843, USA}  
\author{J.T.~White} \affiliation{Texas A \& M University, Department of Physics, College Station, TX 77843, USA}  
\author{T.J.~Whitis} \affiliation{Case Western Reserve University, Department of Physics, 10900 Euclid Ave, Cleveland, OH 44106, USA} \affiliation{SLAC National Accelerator Laboratory, 2575 Sand Hill Road, Menlo Park, CA 94205, USA} \affiliation{Kavli Institute for Particle Astrophysics and Cosmology, Stanford University, 452 Lomita Mall, Stanford, CA 94309, USA}
\author{M.S.~Witherell} \affiliation{Lawrence Berkeley National Laboratory, 1 Cyclotron Rd., Berkeley, CA 94720, USA}  
\author{F.L.H.~Wolfs} \affiliation{University of Rochester, Department of Physics and Astronomy, Rochester, NY 14627, USA}  
\author{D.~Woodward} \affiliation{Pennsylvania State University, Department of Physics, 104 Davey Lab, University Park, PA  16802-6300, USA}  
\author{J.~Xu} \email[Corresponding author, ] {xu12@llnl.gov}\affiliation{Lawrence Livermore National Laboratory, 7000 East Ave., Livermore, CA 94551, USA}  
\author{K.~Yazdani} \affiliation{Imperial College London, High Energy Physics, Blackett Laboratory, London SW7 2BZ, United Kingdom}  
\author{C.~Zhang} \affiliation{University of South Dakota, Department of Physics, 414E Clark St., Vermillion, SD 57069, USA}  

\date{\today}% It is always \today, today,
             %  but any date may be explicitly specified

\begin{abstract}

Various dark matter models predict annual and diurnal modulations of dark matter interaction rates in Earth-based experiments 
as a result of the Earth's motion in the halo. 
Observation of such features can provide generic evidence for detection of dark matter interactions. 
This paper reports a search for both annual and diurnal rate modulations in the LUX dark matter experiment 
using over 20 calendar months of data acquired between 2013 and 2016. 
This search focuses on electron recoil events at low energies, 
where leptophilic dark matter interactions are expected to occur and 
where the DAMA experiment has observed a strong rate modulation for over two decades. 
By using the innermost volume of the LUX detector and developing robust cuts and corrections, 
we obtained a stable event rate of 2.3$\pm$0.2~cpd/\keVee/tonne, 
which is among the lowest in all dark matter experiments. 
No statistically significant annual modulation was observed in energy windows up to 26~\keVee. 
Between 2 and 6~\keVee, this analysis demonstrates the most sensitive annual modulation search up to date, 
with 9.2$\sigma$ tension with the \dl\ result. 
We also report no observation of diurnal modulations above 0.2~cpd/\keVee/tonne amplitude between 2 and 6~\keVee. 

\end{abstract}

%Use showkeys class option if keyword

\maketitle

%\tableofcontents

%%
%% Start line numbering here if you want
%%
%\linenumbers

%% main text

\section{Introduction}
\label{sec:intro} 

Dark matter direct detection experiments search for kinetic energy transfer from hypothetical dark matter particles to target atoms in low background detectors. 
In a variety of dark matter models, 
dark matter-matter interactions may produce recoiling nuclei or electrons at very low energies, 
which then may be detected by state-of-the-art particle detectors. 
Over the past decade, direct detection experiments have greatly improved their sensitivities 
to nuclear recoil (NR) dark matter interactions--the cross section of which is coherently enhanced for the spin-independent channel--but 
no definitive detection has been made up to date~\cite{LUX2016_Run3_4, XENON1T_2017, PandaX2017_DM, CDMS2018_DM}. 
Electron recoil (ER) dark matter interactions~\cite{Kopp2009_LeptDM, Fox2009_LeptDM}, 
on the other hand, 
are relatively less discussed due to the model complexity and the predominant ER background in particle detectors from natural radioactivity. 

A generic feature expected of dark matter interactions is temporal changes of interaction rates in Earth-based detectors. 
Such rate modulations can occur as a result of the relative motion of the Earth in the dark matter halo~\cite{Drukier1986_Modulation}. 
The most widely discussed dark matter rate modulation is an annual modulation due to the Earth orbiting the Sun. 
In a simple picture, the orbital velocity of the Earth adds to that of the solar system in June, 
which can increase the dark matter flux observed by Earthly detectors and also cause a change in the effective  interaction cross section. 
Such effects may lead to a higher overall dark matter interaction rate in June,  
and a lower rate in December~\cite{Drukier1986_Modulation, Lewin_review_1996}. 
The exact amplitude and phase of annual modulations depend on the specific dark matter models, 
and have been formulated in the Weakly Interacting Massive Particle (WIMP) model~\cite{Freese2013_Modulation, Lewin_review_1996}, 
the axion dark matter model~\cite{Semertzidis2015_AxionDMModulation}, 
and dark sector dark matter models, such as mirror dark matter~\cite{Foot2004_DAMAMirrorDM, Foot2013_MirrorDM} 
and two-component plasma dark matter~\cite{Clarke2016_PlasmaDM}. 
Depending on the specific model implementation, 
the interaction signal can be either NRs or ERs in nature.

A controversial dark matter detection claim, by the DAMA experiment (\dn\ and \dl)~\cite{DAMA2013_Phase1, DAMA2018_Phase2},
was made based on the observation of an annual event rate modulation
in a large array of low-background \nai\ detectors deployed at the Gran Sasso underground laboratory. 
Unlike other reported hints of dark matter from \cogent~\cite{CoGeNT2011_Mod}, CDMSII~\cite{CDMS2018_LowMass}, 
and CRESST~\cite{CRESST2012_Excess}, 
the DAMA anomaly has not yet been explained as a background. 
The DAMA modulation signal appears the strongest in an energy window around 3~keV ER equivalent energy (\keVee), 
and vanishes above 6~\keVee, which verifies the stability of the experiment. 
The highest event rate was observed around late May to early June, 
consistent with a dark matter signal. 
Several background hypotheses have been proposed in attempt to explain this signal, 
but none has succeeded in explaining all the modulation features~\cite{DAMA2008_CombinedAnalysis}.  

Although the interpretation of the DAMA modulation signal in a few dark matter models has been 
tightly constrained by other direct detection experiments~\cite{LUX2016_Run3_4, CDMS2018_LowMass,XENON1T_2018, PandaX2017_DM, XENON2015_Leptophilic}, 
a definitive test of DAMA using \nai\ has not been demonstrated as of today. 
On the other hand, searches for dark matter-induced rate modulations 
can offer a generic approach to identify dark matter interactions, 
complementary to the model-driven dark matter searches. 
For ER dark matter models, 
modulation searches also provide a powerful handle to suppress the dominant ER background from natural radioactivity, 
which can be made to be constant through sufficient shielding in deep underground locations. 

The LUX dark matter experiment has achieved one of the highest sensitivities in searching for NR dark matter interactions~\cite{LUX2016_Run3_4}. 
The low-energy ER background rate in LUX is over two orders of magnitude lower than that in \dl, 
and is among the lowest demonstrated in particle detectors. 
This low ER background rate and the multi-year operation of LUX 
make it well suited to search for annual modulation signals from ER dark matter interactions. 
This paper presents a search for such low-energy ER modulations 
using the complete LUX data set~\cite{LUX2016_Run3_4}. 
This analysis focuses on the low energy window of 2--6~\keVee, 
but also extends to higher energies up to 26~\keVee. 

In addition to annual modulation searches, 
we also conducted a search for diurnal rate modulations between 2 and 6~\keVee. 
Diurnal modulations in dark matter interaction rate may be induced by the rotation motion of the Earth around its spin axis, 
with a similar mechanism to that for the annual modulation theories discussed above. 
Due to the lower rotating velocity of the Earth compared to the orbital velocity, 
the diurnal modulation amplitude is usually predicted to be much smaller than that of annual modulations~\cite{Lewin_review_1996}. 
For example, \dl\ estimated the expected diurnal modulation amplitude in their NaI(Tl) detectors 
if the observed signals were due to WIMP dark matter interactions, 
and concluded it is beyond the sensitivity of the \dl\ experiment~\cite{DAMA2014_DiurnalMod}. 
However, for dark sector dark matter models that consider possible interactions between the galactic dark matter wind and Earth-captured dark matter, 
the Earth's spin plays a more significant role in affecting the dark matter flux close to the surface of the Earth, 
which can significantly enhance the relative  amplitude of diurnal modulations~\cite{Clarke2016_PlasmaDM, Foot2015_DiurnalMod}.  
In these dark matter models, 
the diurnal modulation effect could possibly manifest itself in low background experiments like LUX. 

This paper is organized as follows: 
Sec.~\ref{sec:lux} reviews the operation of the LUX dark matter experiment and the observed ER background in the detector; 
Sec.~\ref{sec:analysis} explains the analysis cuts and corrections that we developed to obtain long-term stability in the LUX data set; 
In Sec.~\ref{sec:results}, we present the results of the annual and diurnal modulation searches and discuss the physical implications; 
In Sec.~\ref{sec:concl}, we conclude this work.  

\section{The LUX dark matter experiment}
\label{sec:lux} 

The LUX dark matter detector was located 1480 meters (4850 feet) underground in the Davis Cavern of the Sanford Underground Research Facility (SURF). 
The active LUX detector was a dual-phase xenon time projection chamber (TPC) hosted in a 7.6 m (diameter) by 6.1 m (height) water tank. 
The TPC contained 370~kg of ultra-pure liquid xenon in a titanium cryostat. 
Energy deposited by particle interactions in the liquid xenon induced two measurable signals: 
scintillation photons 
and ionization electrons that escaped electron-ion recombination. 
The former was promptly detected by two arrays of photomultipliers (PMTs), 
one array above the TPC and the other below the TPC.  
For the latter to be detected, 
the ionization electrons were first drifted towards the top of the liquid with an electric field; 
once they entered the thin gas layer above the liquid under the effect of a stronger electric field, 
they produced secondary electroluminescence, 
which was then collected by the PMTs. 
The distribution of the electroluminescence signal was highly localized in the top PMT array, 
enabling the X-Y position of the ionization event to be accurately determined. 
The drift time of the electrons in the liquid, or the time delay between the prompt scintillation (S1) and delayed electroluminescence (S2) signals, 
provided an estimate of the depth of the interaction, 
so the 3-D position of the particle interactions could be reconstructed.  
For more information on the LUX detector, interested readers can refer to \cite{LUX2012_Detector}. 

The complete LUX search for WIMP dark matter consisted of two operation campaigns. 
The first one collected data from April to October 2013, referred to as WS2013 hereafter; 
the second one started in September 2014 and was concluded in May 2016, referred to as WS2014-16. 
These two campaigns covered over 25 calendar months of data collection in total, 
but due to operation interruptions such as calibrations, 
only 20 months' data were suitable for dark matter search analysis. 

The underground location of the LUX experiment reduced the cosmic muon flux by a factor of 10$^{7}$ compared to that at surface. 
As such, background due to direct cosmic rays in the experiment was negligible compared to that from natural radioactivity, 
and the impact from the seasonal fluctuation of cosmic ray flux on the experiment can be ignored. 
The water shielding suppressed environmental gamma-ray and neutron backgrounds by at least 9 orders of magnitude. 
Radon gas background in the water tank was mitigated through constant nitrogen gas purge. 
Due to its large mass and heat capacity, the water tank also functioned as a heat bath to damp any sudden temperature fluctuation in the Davis Cavern. 
Due to a detector warm-up and cool-down cycle from 2013 to 2014, 
the absolute temperature of the liquid xenon shifted from 173~K in WS2013 to 177~K in WS2014-16. 
However, the temperature variation was controlled to be $<$ 0.1~K in the WS2013 data 
and $<$ 0.3~K in the WS2014-16 data used for this analysis. 
Similarly the gas pressure in the detector shifted from 1.58~bar in WS2013 to 1.92~bar in WS2014-16, 
but the pressure was stable at a level of $<$ 0.03~bar for both WS2013 and WS2014-16. 
Despite other changes, 
the liquid level in the detector was kept stable to within $<$ 0.2~mm for the whole operation. 
As will be discussed in Section~\ref{subsec:goldcut}, 
possible changes in the detector performance due to the temperature and pressure shifts between WS2013 and WS2014-16, 
such as that in the S2 gain, 
were calibrated and corrected for in the analysis. 

During WS2013, we observed a possible event rate excess around 3~\keVee\ in the ER energy spectrum, 
at an estimated strength of 1-2~cpd/\keVee/tonne, 
and it was not expected from background models~\cite{LUX2015_Reanalysis, LUX2017_Axion}. 
These events appeared to distribute uniformly in the active xenon volume, 
and they are often attributed to \arsev\ contamination 
-- which is also a possible background in the DAMA experiments~\cite{McKinsey2018_Ar37} --
in the xenon from initial xenon production or air leakage during operations~\cite{LUX2018_Run3PRD}. 
However, no conclusion can be drawn for the origin of these excess events in LUX 
based on measurements of the air leakage rate into LUX and the \arsev\ concentration in the SURF air. 
In WS2014-16, the excess at 3~\keVee\ was determined to be statistically insignificant, 
partially because the field distortion near the detector walls~\cite{LUX2017_3DField} prevented a large fiducial volume from being used in a robust analysis, 
as explained in Section~\ref{subsec:fiducial}. 

This paper studies the temporal behavior of ER events in the LUX detector using data from both WS2013 and WS2014-16, 
searching for both annual modulations and diurnal modulations. 
The primary energy region of interest is below 6~\keVee, 
where \dl\ observed a strong event rate modulation, 
and where such signals are usually discussed in various dark matter models. 
This analysis energy window also covers the energy region for the LUX ER event excess. 
In addition, we extend the annual modulation search up to 26~\keVee. 

\section{Data analysis}
\label{sec:analysis} 

Essential for a sensitive and robust modulation search are a low background event rate and a stable detector operation. 
A low event rate of 3.6~cpd/\keVee/tonne below 5~\keVee\ has been demonstrated in the LUX WIMP search analysis~\cite{LUX2013_FirstResult}, 
and it could be further reduced with more stringent analysis cuts. 
The stability of the LUX experiment, however, was compromised by an evolving electric field problem that resulted from the grid conditioning campaign 
following WS2013~\cite{LUX2017_3DField}. 
As a result, the S1/S2 production and collection in later stages of the LUX experiment differed significantly from WS2013, 
and continued to deteriorate throughout WS2014-16. 
This section discusses the cuts and corrections that were developed to restore stability in the LUX data. 

\subsection{Fiducial cut}
\label{subsec:fiducial}

\begin{figure}[h!]
\centering
\includegraphics[height=.37\textwidth]{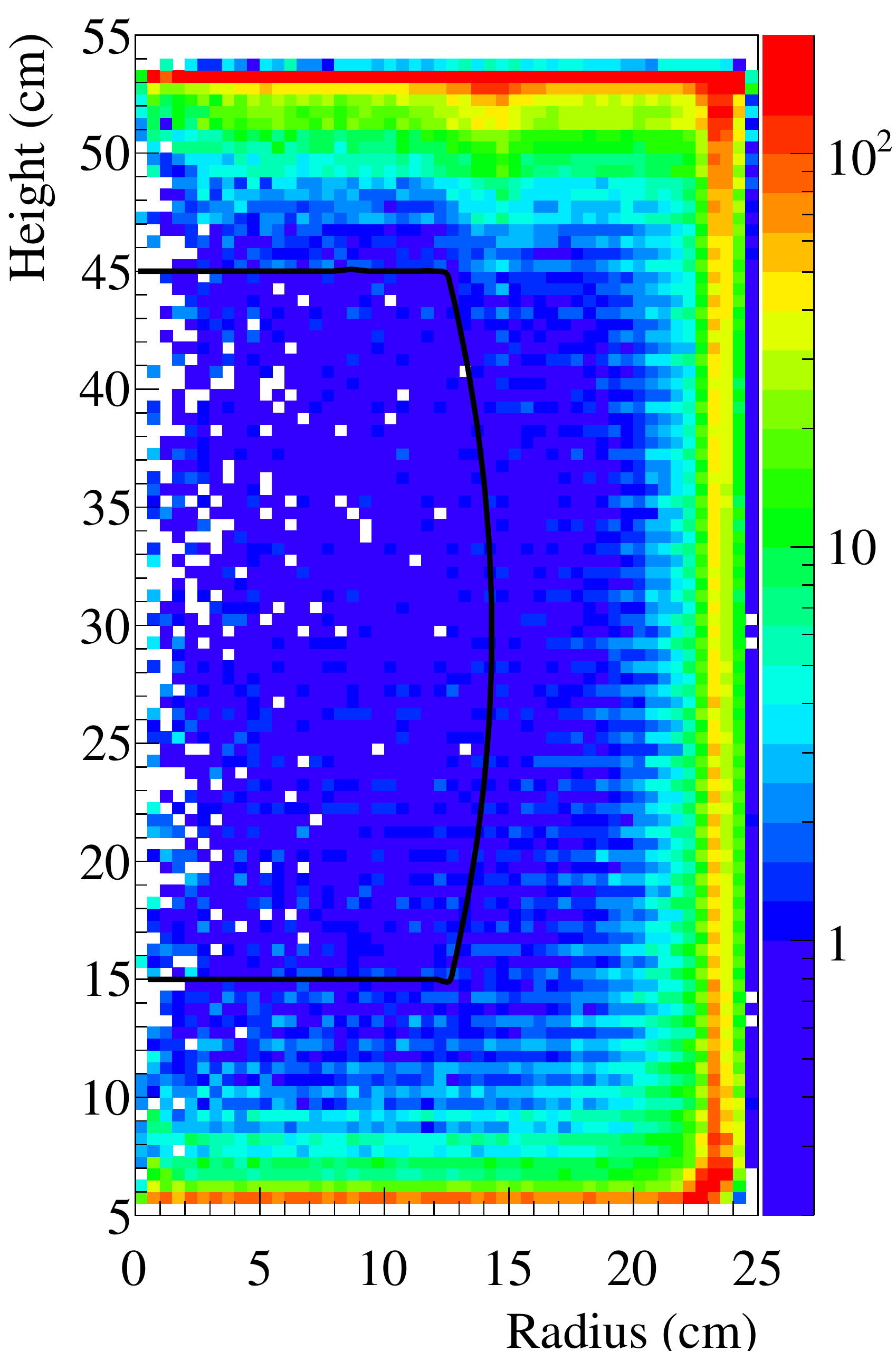}
\includegraphics[height=.37\textwidth]{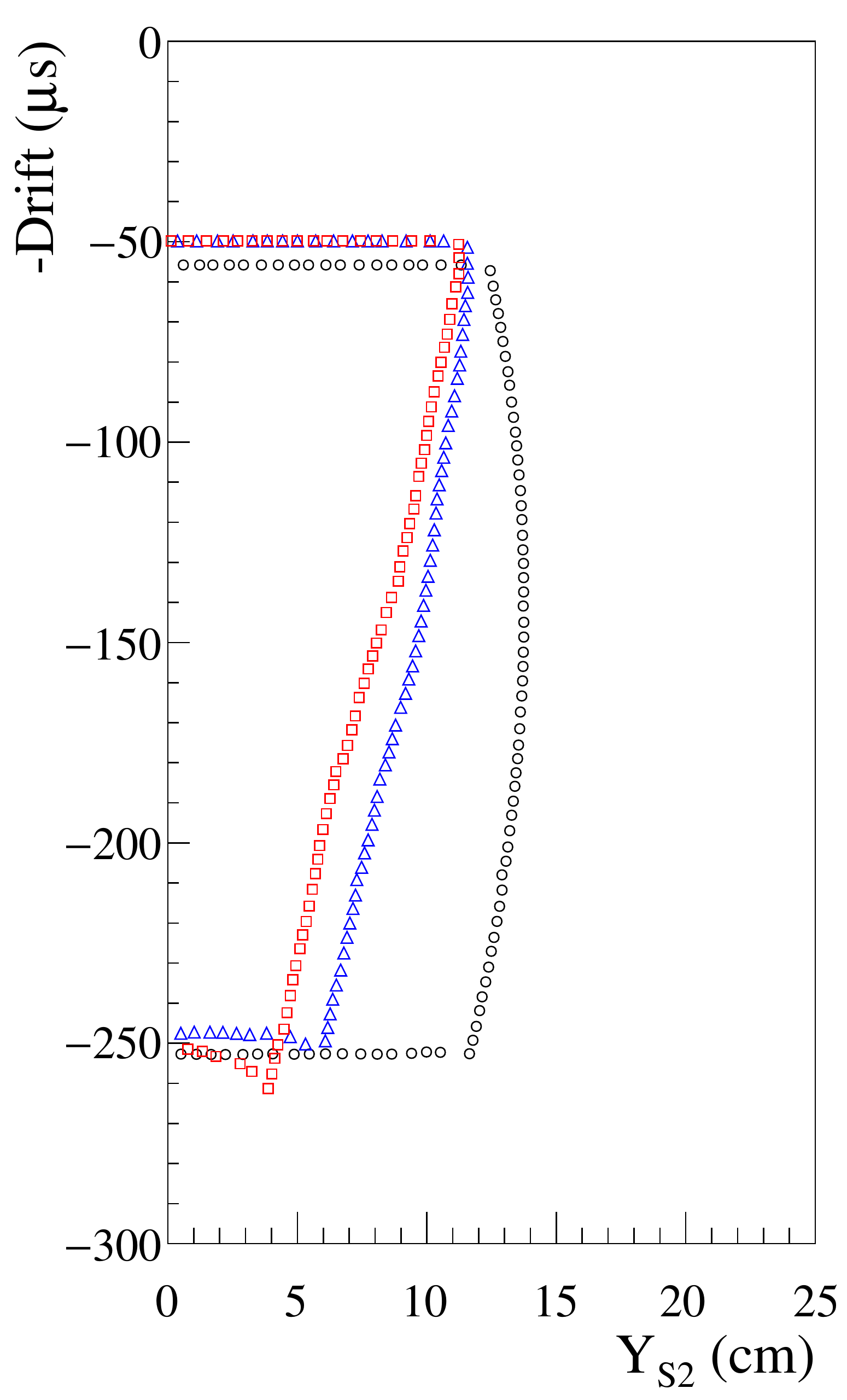}
\caption{{\bf Left:} Illustration of the FV used in this analysis (black line) in comparison to the density distribution of single scattering events ($<$500~\keVee) in WS2013; 
the coordinates used are estimated real-world positions corrected based on the S2 positions and the simulated electric field. 
{\bf Right:} Illustration of the fiducial cut applied to the drift time and S2 positions along $x_{S2}$=0 at different times, 
including WS2013 (black circles), early (blue triangles) and late (red squares) WS2014-16. 
This FV is $\sim$2-3 times smaller than that used in the LUX WIMP searches~\cite{LUX2015_Reanalysis, LUX2016_Run3_4}. }
\label{fig:fv}
\end{figure}

The underground location and the water shielding reduced the background event rate in LUX drastically. 
Remaining background in LUX was dominated by gamma rays from the detector components in proximity to the active volume, 
and by alpha-decays on the Polytetrafluoroethylene (PTFE) reflector surface that surrounded the liquid xenon. 
Thanks to the strong self-attenuation power of liquid xenon and the excellent position reconstruction capability of LUX~\cite{LUX2018_PositionRecon}, 
most of these background events were identified to be near the edge of the active volume, 
as illustrated in Figure~\ref{fig:fv} (left) 
and can be rejected from the dark matter analysis. 

However, due to the electric field distortion in WS2014-16, the observed event positions were biased towards the center of the TPC, 
especially for those close to the bottom of the liquid xenon volume. 
This behavior caused both a position bias and an inhomogeneous position resolution, 
both of which deteriorated over time. 
Therefore, a simple fiducial cut applied to the observed event positions, namely $x_{S2}$, $y_{S2}$ and the drift time, 
would correspond to a time-dependent physical fiducial volume (FV), 
and thus produce a background rate varying with time. 
To address this problem, the FV was defined in the real-world space, 
and then a position map -- which was derived from a dedicated 3D electric field study~\cite{LUX2017_3DField} --  
was used to map the fiducial boundary to the reconstructed S2 space before comparing with event positions. 

The FV in the real space was defined as radially symmetric. 
The radial boundary at each depth was chosen to ensure that within this boundary 
the low-energy ER background rate did not vary significantly with radius or azimuth angle~\footnote{To avoid bias, 
only the ER data between 6~\keVee\ and 26~\keVee\ were used to determine the FV boundary, excluding the energy region of interest below 6~\keVee. }. 
Because of the deterioration of the electric field over time, 
a small FV was chosen to make sure the above criterion is met even for the worst field distortion. 
The final corrected FV had a maximum radius of 14~cm in the center, 
and the value decreases towards the top and bottom of the TPC, as illustrated in Figure~\ref{fig:fv} (left). 
The top and bottom limits of this FV were chosen to be 9.2~cm above the cathode grid and 8.8~cm below the liquid surface, 
following a similar criterion as explained for the radial limits. 
The same FV in the observed position space is illustrated in Figure~\ref{fig:fv} (right), 
which shows very significant time dependence. 

\begin{figure}[h!]
\centering
\includegraphics[width=.46\textwidth]{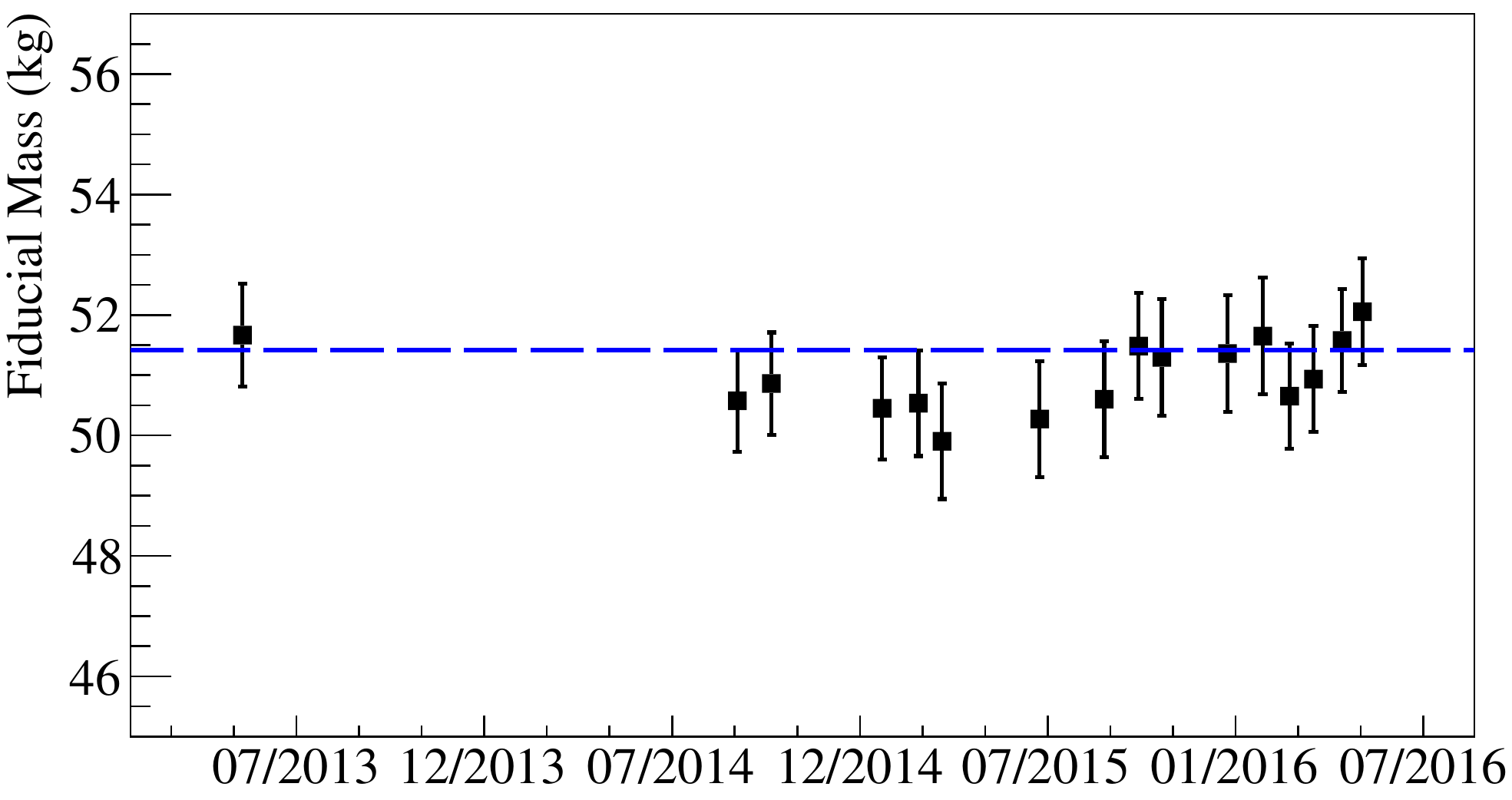}
\caption{The fiducial mass calculated from the FV geometry (blue dashed line) and from \krthree\ calibration data (black squares with error bars). The error includes uncertainties from the \krthree\ event selection criteria, from the total active mass, and from the field map interpolations. }
\label{fig:fm}
\end{figure}

The fiducial mass was estimated from two independent approaches. 
The first method is a direct calculation using the geometry of the FV and a xenon density of 2.9~kg/cm$^3$ at 175~K, 
yielding a mass of 51.4~kg. 
The second method is an indirect estimation based on the total active xenon mass in LUX and the fraction taken by the FV. 
In the LUX experiment, \krthree\ gas was regularly introduced into the detector for position and energy calibrations.  
\krthree\ decay events have been shown to distribute uniformly in the active volume several minutes after the source injection~\cite{LUX2017_Kr83m}. 
Therefore, the fraction of \krthree\ events in the FV serves as a good indication of the fraction of volume accepted by the fiducial cut. 
Figure~\ref{fig:fm} shows the estimated fiducial mass for 16 \krthree\ calibration data sets 
over the course of 3 calendar years. 
Despite the evolving electric field problem, the fiducial mass was stable at a level of 2\%.

With the stringent fiducial cuts to choose only events in the very center of the LUX detector, 
this fiducial mass is substantially smaller than that used in previous LUX analyses, 
145.4~kg in \cite{LUX2015_Reanalysis} and 98.4-107.2~kg in \cite{LUX2016_Run3_4}. 
In addition to helping restore long-term data stability in the analysis, 
this choice of FV also significantly reduced various background at the edge of the active volume. 
These background sources include low-energy external gamma rays, 
decays of radon progeny on PTFE surfaces, 
and mostly importantly, 
the L-shell electron capture decays of \xesev, 
which can produce an ER background in the signal region of interest for this modulation analysis. 
\xesev\ can be produced at trace levels when xenon is exposed to cosmic rays at surface, 
and has been observed in the LUX detector~\cite{LUX2017_Xe127}. 
Particularly, the L-shell electron capture decays of \xesev\ can produce a peak at 5.2~keV 
when the accompanying gamma rays (dominantly 203~keV) escape the active xenon volume. 
However, with this chosen small FV in the center of LUX, 
very few 203~keV gamma rays originating from the FV can escape. 
As a result, this 5.2~keV background is substantially reduced.  
This background was evaluated to be negligible in this analysis 
after the data acquired in early WS2013 were excluded, 
as discussed in Section~\ref{subsec:exclcut}. 

\subsection{Single scatter cut}
\label{subsec:goldcut}

Another powerful handle to reject background is the single scatter (SS) requirement. 
The chance of a dark matter particle scattering more than once within the LUX detector is vanishingly small, 
while gamma-ray and neutron background could produce multiple interaction vertices in $\sim$50~cm of liquid xenon. 
For an event to be considered for this analysis, it is required to have only one valid S1-S2 pair in the 1~ms data acquisition window. 
A valid S1 signal is defined as a fast pulse (10s of ns) in which at least 2 or more PMTs each recorded one or more detected photons (phd). 
A valid S2 pulse is defined as a wide pulse (a few $\mu$s) with a characteristic rise and fall time. 
The SS cut selects events with a single S2 pulse in the whole event window and a single S1 pulse before the sole S2 pulse. 

Several factors could impact the efficiency of the SS cut, 
especially at low energies where the S1s consisted of only a few photons. 
Any changes in the optical properties of detector components, 
such as the PTFE reflectivity or the liquid level in the top of the TPC, 
could cause the light collection efficiencies for both S1s and S2s to vary with time. 
Changes in the liquid level, in the gas pressure or in the detector temperature, 
can further modify the production efficiency of S2 electroluminescence signals. 
The evolving electric field in LUX is also expected to introduce time dependence in 
both the production efficiency and the collection efficiency of S1s and S2s. 

In the LUX experiment, a wide range of techniques were developed to measure the detection efficiencies for S1s and S2s, 
abbreviated as $g1$ and $g2$, respectively. 
$g1$ is defined simply as the fraction of S1 scintillation light that was collected by the PMTs; 
$g2$ is defined as the number of photons detected for every primary electron produced in the liquid, 
and it includes contributions from the electron extraction efficiency, the electroluminescence production efficiency and the S2 light collection efficiency. 
With the $g1$ and $g2$ corrections, 
the overall energy of an event can be reliably estimated as 
$E=W(\frac{S1_c}{g1}+\frac{S2_c}{g2})$, 
where $W$=13.7 eV is the average energy required to produce either one ionization electron or one scintillation photon in liquid xenon~\cite{DahlThesis}, 
and S1$_c$ and S2$_c$ are the position-corrected energy variables. 
Throughout the LUX experiment, $g1$ and $g2$ values were regularly monitored through internal and external calibrations, 
including \krthree~\cite{LUX2017_Kr83m}, \hthree~\cite{LUX2016_3H}, and xenon activation lines following neutron calibrations~\cite{LUX2017_Yield}. 
The values of $g1$ and $g2$ remained stable within WS2013, 
and the drift was estimated to be $<$8\% from the beginning of WS2014-16 to the end. 
By defining $g1$ and $g2$ as empirical functions of time, 
the effects of small changes in the detector operation parameters, such as the liquid level, liquid temperature and gas pressure, 
were corrected for in the data. 

To evaluate the SS cut efficiency with corrected $g1$ and $g2$ values, 
an all data-driven approach was used based on the \hthree\ calibration data, as outlined in \cite{LUX2016_3H}. 
\hthree\ radioactivity was regularly introduced into the LUX detector to calibrate low energy ER events. 
The \hthree\ beta spectrum has an end point energy of 18.6~keV, 
with a peak at 2.5~keV and a mean energy of 5.6~keV, 
making it ideal for efficiency studies in our energy region of interest. 
The spectral shape of \hthree\ beta decays is well known both theoretically and experimentally, 
allowing the SS \hthree\ data in the FV to be fitted to the known \hthree\ spectrum at the high energy end, 
where the acceptance of the SS cut is $\sim$100\%. 
Then the fitted \hthree\ spectrum (with 100\% efficiency) was extrapolated to low energies 
and compared to the observed event spectrum, 
for the relative cut acceptance to be calculated as a function of energy. 

\begin{figure}[h!]
\centering
\includegraphics[width=.43\textwidth]{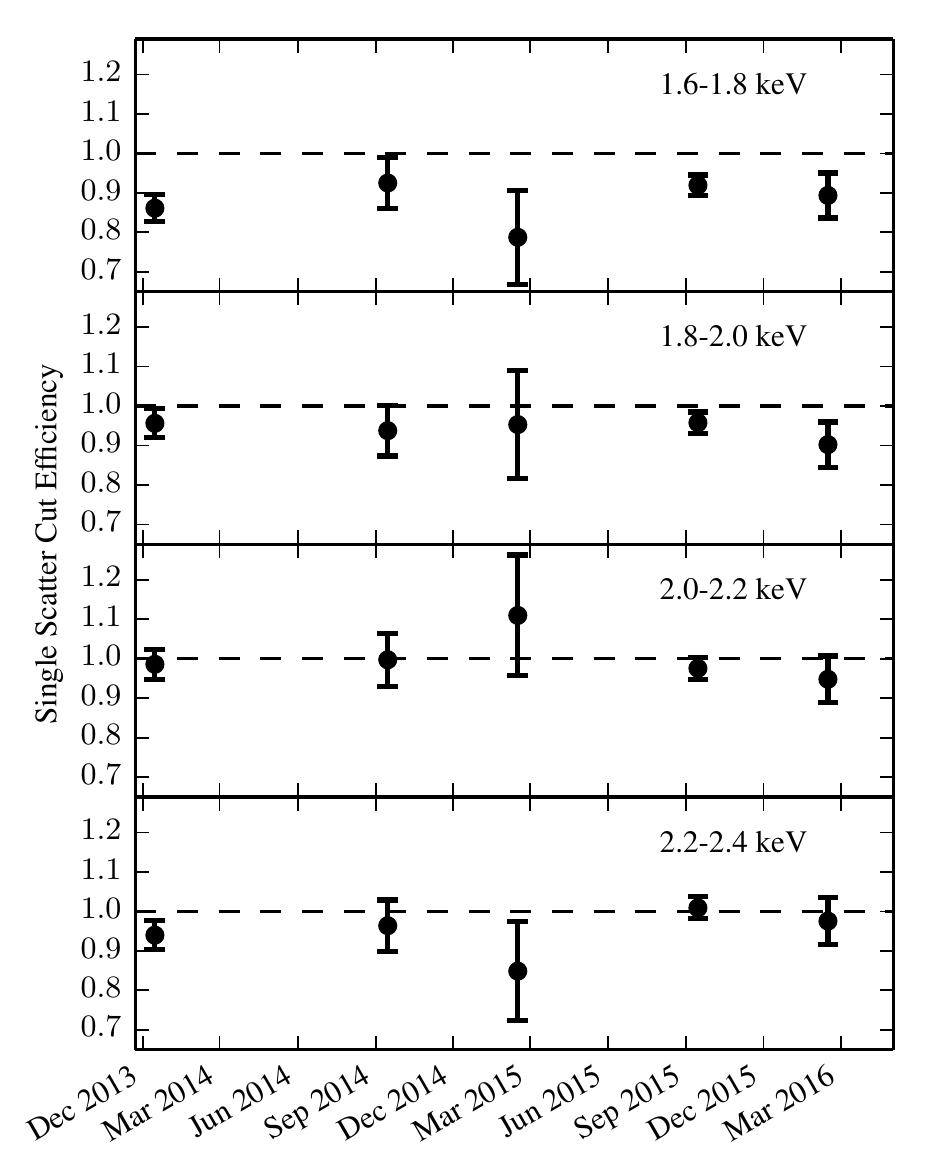}
\caption{The efficiency for the single scatter cut as a function of time  evaluated for events between 1.6~\keVee and 2.4~\keVee. 
The efficiencies were calculated from \hthree\ data taken in December 2013, in September
2014, in February 2015, in September 2015 and in February 2016.} 
\label{fig:goldeneff}
\end{figure}

Fig.~\ref{fig:goldeneff} shows the SS cut efficiency evaluated for events between 1.6~\keVee\ and 2.4~\keVee\ 
as a function of time. 
The efficiencies were calculated for \hthree\ data acquired in December 2013, 
in September 2014, in February 2015, in September 2015 and in February 2016. 
Thanks to the $g1$ and $g2$ corrections, 
the observed efficiencies are mostly stable over time, especially for events above 1.8~\keVee. 
The low energy analysis threshold was conservatively chosen to be 2~\keVee\ in the modulation analysis.  
In the main analysis energy window of 2-6~\keVee, the SS cut efficiency is mostly consistent with 100\% and remains stable at a 5\% level. 
Results of this efficiency study were also confirmed at 1$\sigma$ level with independent simulations using the NEST package~\cite{NEST2013}, 
with the evaluated electric field taken as an input. 

As will be discussed in Section ~\ref{subsec:anmod}, 
at higher energies the SS efficiency decreases slightly due to misidentified S2 pulses. 
However, this small drift is not expected to significantly impact the analysis. 

\subsection{Data quality cuts}
\label{subsec:qualcut}

In principle, the fiducial cut and the single scatter cut can provide sufficient background rejection for this analysis. 
However, uncertainties in the pulse finding and pulse classification algorithms 
can make certain background events appear as single scatter events in the FV. 
The most relevant background of this kind is randomly paired S2 pulses and S1 pulses (or S1-like pulses) during high pulse rate periods. 
This section discusses the main data quality cuts that were developed to suppress such background events. 

It was observed in LUX that the rate of small S2s and single electrons increased significantly in periods after high energy events.  
Due to the high rate, small S2 pulses may be paired with S1 pulses, 
or mis-tagged S1 pulses, 
and then incorrectly identified as single scatter events. 
In this analysis, we applied a 20~ms veto after each event that had a total pulse integral larger than 10$^5$~phd ($\sim$ 300~\keVee). 
In addition, we also applied a 20~ms veto cut every time the data acquisition system went inactive for $>$ 3~ms, 
in case a high energy event occurred in this window but was not recorded. 
The loss of live-time due to this veto cut was calculated to be $\sim$10\%. 

A similar background can rise in the same event window of a high energy event 
when the large S2 pulses were distorted and failed to be identified by the pulse classification algorithm. 
In this situation, small S2 pulses right after the large S2s may be mis-paired with S1-like pulses, producing a false single scatter event. 
To reject such background events, we require the identified S1 and S2 pulse pair in a single scatter event to contain more pulse area than the unaccounted-for pulse area in the same event window. 
We do not expect any significant loss of physical single scatter events from this cut in the energy region of interest. 

\begin{figure}[h!]
\centering
\includegraphics[width=.4\textwidth]{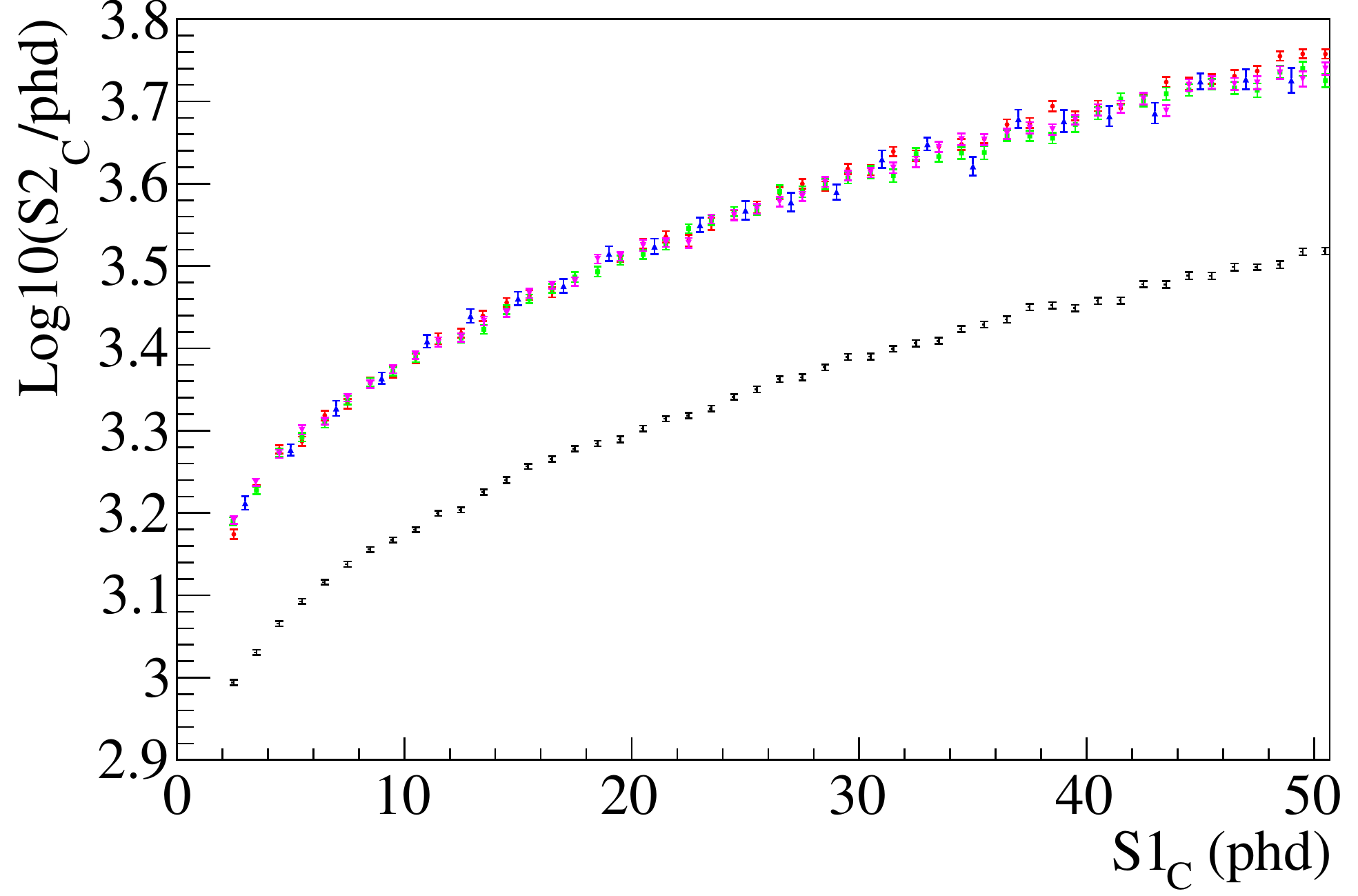}
\caption{The central distribution of ER events in the S1-log$_{10}$(S2) (position corrected) parameter space, derived from \hthree\ data acquired in December 2013 (black dots), September 2014 (green squares), February 2015 (blue triangles), February 2016 (magenta downwards triangles), and \cfour\ data in July 2016 (red circles). 
Data from WS2014-16 exhibited higher S2 values than WS2013 because of improved high voltage; 
all WS2014-16 data in the FV were consistent with each other despite the evolving field distortion at large radii. 
S1s of 50~phd approximately correspond to 11~\keVee. }
\label{fig:erband}
\end{figure}

A small fraction of mis-paired background events survived both the veto and the pulse area fraction cut. 
These events were rejected using an ER identification cut 
because ER events follow a certain S1-S2 distribution governed by the energy partition between scintillation and ionization channels, 
but randomly paired events do not. 
The exclusion of the NRs with this cut is not expected to bias this search because 
previous LUX analysis~\cite{LUX2016_Run3_4} has concluded no observation of NR event excess in the same data set as used here. 
In addition, dark matter-NR interactions are generally suppressed in leptophilic dark matter models~\cite{Kopp2009_LeptDM}. 
The ER acceptance region in the S1-S2 parameter space was defined using data from \hthree\ calibrations and \cfour\ calibration, 
which produced pure low energy ERs with high statistics. 
During WS2014-16, \hthree\ was injected into LUX approximately every 6 months 
to monitor possible changes in the ER band in the evolving electric field. 
Thanks to the choice of a small FV in the center of LUX, 
where the field distortion was the smallest, 
the measured ER band did not change significantly in WS2014-16, as shown in Figure~\ref{fig:erband}. 
In this analysis, events that were $>$3 standard deviations from the mean of the ER band were rejected, 
resulting in a time-independent ER event acceptance of 99.7\%. 
It has been reported that the ER events induced by Compton scattering may exhibit a slightly different S1-S2 energy partition from that of pure beta decays like \hthree\ and \cfour~\cite{neriX2017_ERCompton}, 
but due to the conservatively chosen $\pm3\sigma$ acceptance region, 
the impact of a slight ER band shift on the signal acceptance is at the sub-percent level. 

During WS2014-16, NR-like events were artificially assembled and injected into the LUX data stream 
as a means to calibrate the WIMP dark matter analysis. 
Due to the overlap between the NR distribution and that of ERs, 
the artificially introduced dark matter events were excluded from this analysis. 

\subsection{Live-time exclusions}
\label{subsec:exclcut}

As mentioned above, controlled radioactivities were regularly introduced into the LUX detector to calibrate its performance. 
To ensure the background rate stability, a significant fraction of LUX data during and after calibrations were excluded. 

\krthree\ sources were injected into LUX weekly, 
and we excluded the \krthree-dominated data sets from this analysis, 
starting from 1 hour before the source injection until the \krthree\ decay rate dropped to $<$5\% of the background event rate in the \krthree\ energy region. 
On average, each \krthree\ calibration resulted in $\sim$24-30 hours of dead time. 
The residual \krthree\ contamination is not expected to produce a background in the energy region of interest for this analysis, 
owing to the isomeric transition decays of \krthree\ with a decay energy of 41.6~keV. 
For neutron calibrations that could activate short-lived isotopes in and around the LUX TPC, 
we excluded 2 weeks of data following extensive deuterium-deuterium neutron calibration campaigns~\cite{LUX2016_DDCalibration}, 
and 2 days after short AmBe calibrations and \cftwo\ calibrations. 
\hthree\ has a half-life of 12 years, but the compound carrying radioactive \hthree\ (CH$_4$) can be removed by the getter that purified the xenon continuously. 
As a result, the detected \hthree\ rate was observed to decay with a half-life of 6 hours according to \cite{LUX2016_3H}. 
Therefore, only 4 days of data following each \hthree\ injection were excluded. 
A low level of background from initial contamination of cosmogenic \xesev\ radioactivity~\cite{LUX2017_Xe127} and possibly \arsev\ decays in the xenon 
was observed in early LUX data, 
so the first month of WIMP search data in WS2013 were excluded. 
As a result, all data used in this analysis were acquired after the xenon was brought underground for over 4 months, 
or $>$4 half-lives for both \xesev\ and \arsev. 
The residual contamination in the 2-6~\keVee\ window was estimated to be less than 3 events. 

Some detector operations may cause the experimental conditions to change temporarily, 
and we excluded periods when anomalies were observed in the detector temperature, pressure or liquid level. 
Data sets that measured low liquid xenon purity values were also excluded from this analysis. 
The data acquisition system of LUX did not keep track of the change of daylight saving time (DST), 
which was corrected for in this analysis, 
but ambiguity in the event time still occurred in early November. 
As a result, up to 6 hours of data were removed when there was a DST change, 
ensuring no ambiguity for the longest data sets acquired around this time. 

In addition to the large scale live-time exclusions, 
the LUX experiment also excluded live-time segments at much smaller time scales. 
The LUX trigger system implemented a hold off after each acknowledged trigger, 
and the value was set to be 4~ms in early WS2013 and was reduced to 1~ms later~\cite{LUX2016_Trigger}. 
In addition, if a trigger occurred within 500 $\mu$s before the data acquisition is deactivated, 
the recorded waveform may be incomplete; 
these triggers were therefore excluded from the analysis. 

All of the exclusions discussed above were taken into consideration when the effective live-time was calculated for this analysis, 
and the calculation also addressed the situation when two or more exclusions were not mutually exclusive. 
The total remaining live-time was evaluated to be 271 days, 
in comparison to the overall live-time of 427 used in the standard LUX WIMP analysis~\cite{LUX2016_Run3_4}. 

\section{Results and discussions}
\label{sec:results} 

\begin{figure}[h!]
\centering
\includegraphics[width=.43\textwidth]{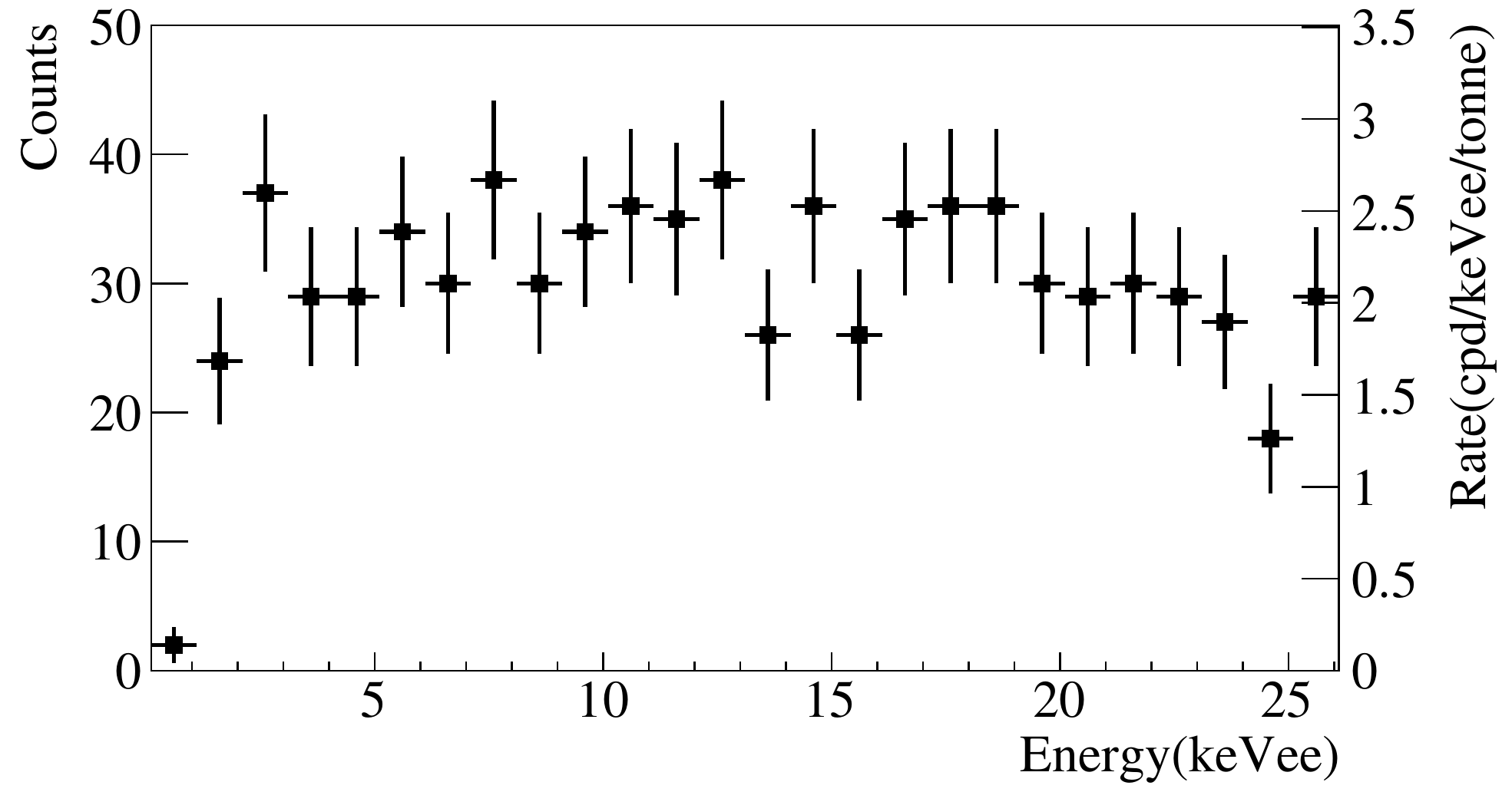}
\caption{The energy spectrum of single scatter ER events in the central 51.4~kg FV of the LUX detector (277 live days total). 
Both the absolute event counts (axis on the left) and the normalized event rates (axis on the right) are shown. 
Due to the stringent live-time exclusion criteria, only 32 live days of data from WS2013 was used, 
and the analysis data set is dominated by WS2014-16 where the 3~\keVee\ event rate excess is less statistically significant. 
}
\label{fig:erspect}
\end{figure}

The combined energy spectrum of single scatter ER events in the central 51.4~kg FV of LUX is shown in Figure~\ref{fig:erspect}. 
The spectral shape is mostly flat in this energy region, 
where the residual events were dominated by Compton scattering of high energy gamma rays 
and by beta decays in the liquid xenon. 
The average ER event rate below 10~\keVee was calculated to be 2.3$\pm$0.21~cpd/\keVee/tonne. 
This rate is significantly lower than that demonstrated in previous LUX analyses \cite{LUX2013_FirstResult, LUX2016_Run3_4, LUX2017_Axion} thanks to the stringent fiducial cut, 
and is among the lowest ever demonstrated in dark matter detectors. 

This section focuses on the searches for annual and diurnal rate modulations in the event rate between 2~\keVee\ and 6~\keVee. 
To estimate possible systematic uncertainties that may not be fully addressed by the methods discussed above, 
we selected the energy window of 6-10~\keVee, 
where the event rate can be mostly explained by background models, 
as a control region. 
For the case of annual modulation search, 
we also extend the analysis for ER events up to 26~\keVee. 
Due to the large number of free parameters in typical ER dark matter interaction models, 
we do not interpret the search result in any specific dark matter models, 
but rather present it as model-independent. 

\onecolumngrid

\begin{figure}[h!]
\centering
\includegraphics[width=.8\textwidth]{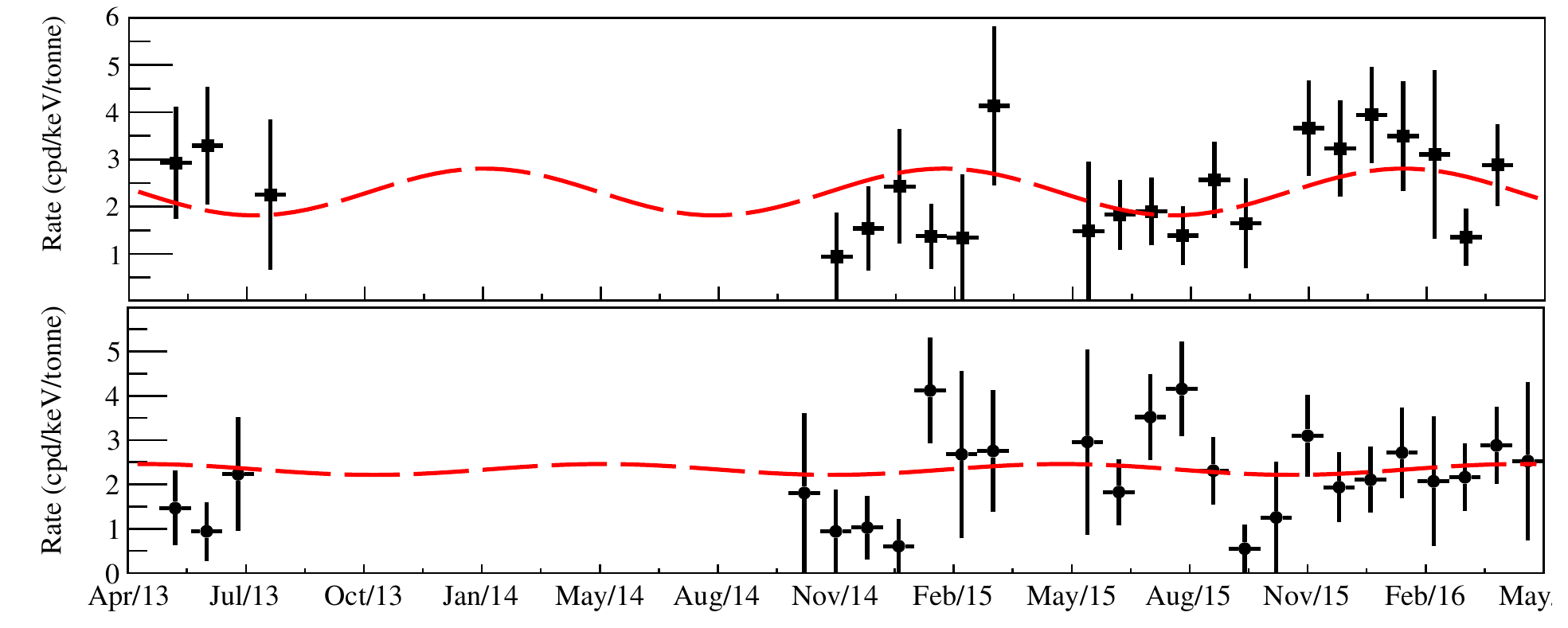}
\caption{The observed LUX event rates in the 2-6~\keVee\ energy window (top) and that in the 6-10~\keVee\ window (bottom) from 2013 to 2016. 
No data exist between 12/2013 and 9/2014 because of detector maintenance. 
Dashed lines illustrate the best fits to an annual modulation model,  
determined using the unbinned maximum likelihood method. 
For the purpose of illustration the live-time is folded in the event rate rather than in the fit function; 
a bin size of 25 days is used. }
\label{fig:anmodrate}
\end{figure}

\twocolumngrid

\subsection{Annual modulation}
\label{subsec:anmod}

With all the aforementioned cuts and corrections applied, 
the time-dependent event rates in the signal region (2-6~\keVee) and in the control region (6-10~\keVee) 
from 2013 to 2016 are shown in figure~\ref{fig:anmodrate}. 
The gap from late 2013 to 2014 was due to detector maintenance between WS2013 and WS2014-16. 
No significant event rate excess around late May to early June as that observed by \dl~\cite{DAMA2013_Phase1} and XENON100~\cite{XENON2017_Modulation}, 
is observed in either group of data. 
Also shown in figure~\ref{fig:anmodrate} are the best fit annual modulation functions to the data, defined as
\begin{equation}
  R(t) = [ A\, \text{cos}(\frac{2\pi}{T}(t-P)) + B ] \times f_{LT}(t)
\end{equation}
where $A$, $T$, and $P$ are the modulation amplitude, 
period (fixed at 1 year), and peak time (days since January 1st), respectively;  
$B$ represents the summed rate of background events and hypothetical dark matter interactions that do not modulate, 
and $f_{LT}(t)$ is the ratio of experimental live-time after all exclusions to time elapsed. 
With the stringent exclusion criteria, 
we expect the residual background event rate to not have a significant time dependence 
and modeled $B$ as a constant. 
As explained in Section~\ref{sec:analysis}, 
background rejection in this analysis was predominantly achieved using exclusions, 
and the loss of exposure is accounted for by $f_{LT}(t)$. 
Further, the explicit cuts (SS cut and ER selection cut) were designed to be conservative 
so the efficiencies are close to 100\%. 
Therefore, no further loss of signal acceptance is assumed in the rest of this analysis, 
and the possible bias on the evaluated modulation amplitude should be $<$5\%. 

\begin{figure}[h!]
\centering
\includegraphics[width=.45\textwidth]{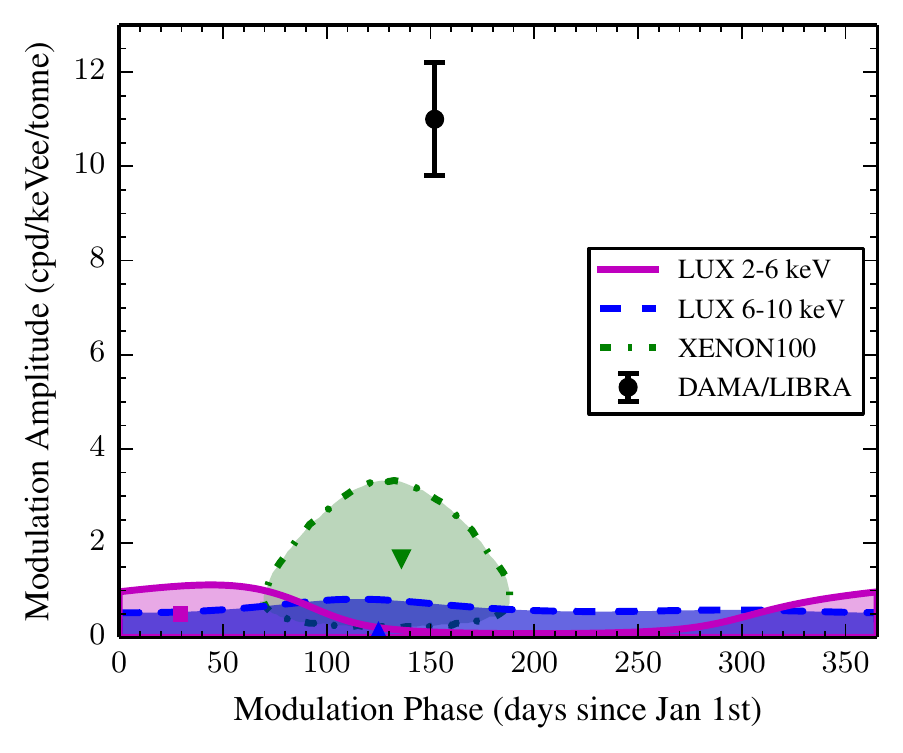}
\caption{The evaluated 90\% LUX contours for the modulation parameters 
in the signal region of 2-6~\keVee (solid line, purple-filled), 
  and that in the control region of 6-10~\keVee (dashed line, blue-filled). 
The DAMA result (\dn\ and \dl) for 2-6~\keVee~\cite{DAMA2013_Phase1} (black dot with error bars) 
and the XENON100 result for 2-5.8~\keVee~\cite{XENON2017_Modulation} (dotted line, green-filled) are also shown for comparison. }
\label{fig:anmodcont}
\end{figure}

Although figure~\ref{fig:anmodrate} shows the data in a binned format, 
the fits were carried out using the unbinned maximum likelihood (UML) algorithm to avoid bias from the binning. 
The log likelihood function in the fits was defined as
\begin{equation}
-\text{ln}(\mathcal{L}) = \int_{T_0}^{T_1} R(t) dt - \sum_{i} \text{ln} R(t_{i})
\end{equation}
where $T_0$ and $T_1$ are the start and end time of the experimental search, 
and $t_i$ represents the detection time of each ER event passing all the cuts. 
The best-fit modulation amplitude was determined to be 0.50~cpd/\keVee/tonne for the signal region with a phase of 30 days, 
and 0.12~cpd/\keVee/tonne with a phase of 124 days for the control region, as shown in figure~\ref{fig:anmodcont}. 

To determine the goodness of the fits, 
the Monte Carlo method was used to generate toy experiments for every combination of test parameters ($A$, $P$). 
In the simulations, the non-modulating event rate was set to be the average rate measured, 
which was also allowed to fluctuate with a Poissonian spread between different data sets simulated. 
For each simulated data set, two UML fits were attempted, 
with one constraining the modulation parameters at the true values, 
and the other with no constraints to search for the global maximum of the UML. 
The test statistic was then defined as the log ratio of the two likelihoods: 
\begin{equation}
q = - \text{ln} \lambda = 
- \text{ln} \frac{\mathcal{L}(\hat{B} | A, P, \{t_i\})}{\mathcal{L}(\hat{A}, \hat{P}, \hat{B} | \{t_i\})}
\end{equation}
where parameters with the ``hat'' symbol represent the values at the maximum (conditional) likelihood. 

The distribution of the test statistic $q$, obtained from Monte Carlo simulations, 
was used as a reference to determine the confidence levels (CLs) of the fits in LUX data. 
The evaluated 90\% confidence region (statistics-only) for the signal region of 2-6~\keVee\ is shown 
in figure~\ref{fig:anmodcont} (solid line, purple). 
This result is consistent with that obtained from using the Wilks Theorem, 
in which -2ln($\lambda$) is approximated as a $\chi^2$ distribution. 
The 90\% CL region covers zero modulation amplitude for all possible phases, 
and does not show any significant increase around 152 days, in contrast to \dl~\cite{DAMA2013_Phase1} and XENON100~\cite{XENON2017_Modulation}. 
Figure~\ref{fig:anmodcont} also shows the 90\% CL region for the control data between 6 and 10~\keVee\ (dash line, blue), 
which remains flat for almost all phases. 
Therefore, we deem any remaining systematic effects, which have not been accounted for in the corrections discussed above, 
to be subdominant, and only focus on the statistical uncertainty in this analysis. 

Thanks to the low ER background rate in LUX and the robust correction algorithms, 
the LUX experiment demonstrates the most sensitive annual modulation search with ER events to date. 
The highest modulation amplitude in the 90\% CL limit is at the level of 1.1~cpd/\keVee/tonne at a phase of 50 days. 
This LUX result is approximately an order of magnitude more sensitive than that of \dl\ 
and a factor of $\sim$3 improvement from XENON100~\cite{XENON2017_Modulation}. 
For a direct comparison with \dl, 
the modulation amplitude was evaluated with the modulation phase fixed at June 2nd (152 days from January 1st). 
In this scenario, we obtained a modulation amplitude of -0.33$\pm$0.27~cpd/\keVee/tonne for the signal region, 
and 0.10$\pm$0.29~cpd/\keVee/tonne for the control region. 
We comment that a negative modulation amplitude corresponds to a modulation that is 180 degrees out of phase, 
and thus is physical. 
The negative portions of the significance contours are not shown in Figure~\ref{fig:anmodcont}, 
but can be inferred by the limit values at 180 degrees phase difference. 
This LUX result is in 9.2$\sigma$ tension with the combined \dl\ and \dn\ result of 11.0$\pm$1.2~cpd/keV/tonne in the same energy window, 
consisting of the most stringent test of \dl\ with any target materials to date. 
The most recent XMASS modulation search reported an energy-dependent 90\% CL limit of 1.3 - 3.2~cpd/\keVee/tonne 
between 2 and 6~\keVee\ at the phase of 152 days~\cite{XMASS2018_Modulation}, 
significantly higher than this LUX result. 

\begin{figure}[h!]
\centering
\includegraphics[width=.44\textwidth]{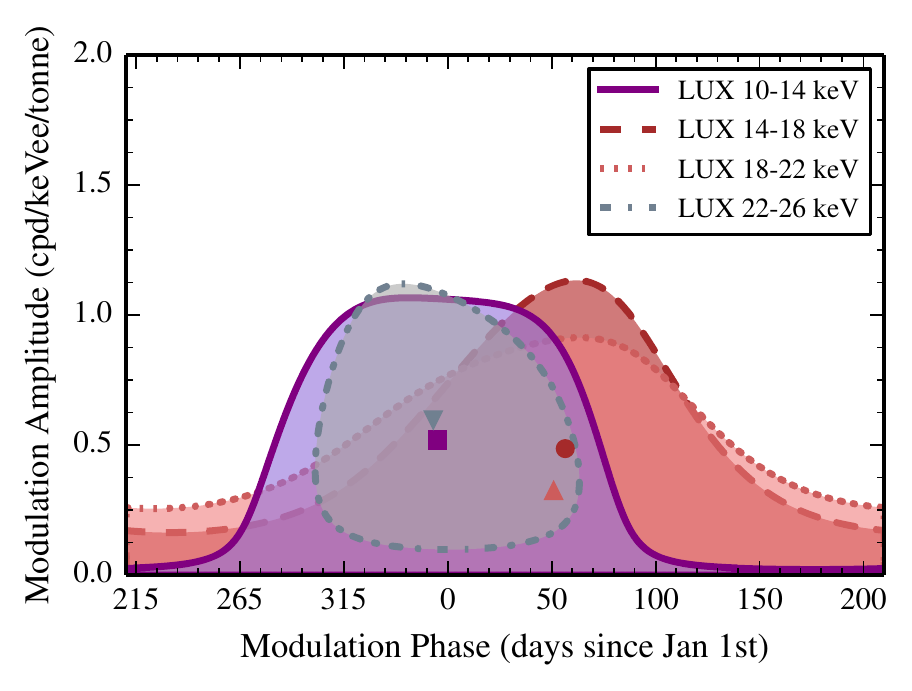}
\includegraphics[width=.44\textwidth]{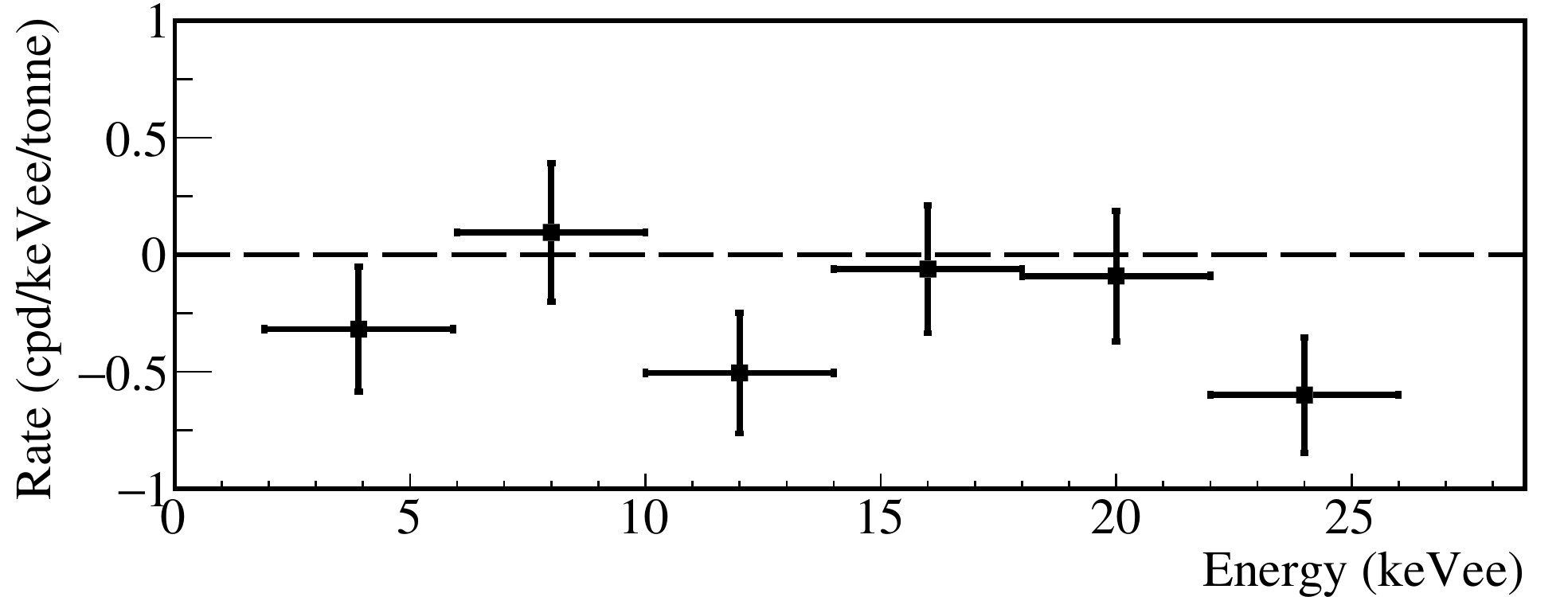}
\caption{
{\bf Top:} The 90\% significance contours in the modulation parameter space for LUX ER events between 10 and 26~\keVee. 
Data is grouped using the same 4~\keVee\ bin size as used in the low energy analysis: 
10-14~\keVee\ (solid, purple), 14-18~\keVee\ (dashed, dark red), 18-22~\keVee\ (dotted, pink), 22-26~\keVee\ (dot-dashed, grey). 
{\bf Bottom:} The best fit modulation amplitude in all LUX ER data below 26~\keVee. 
The modulation phase was fixed to be 152 days in the fits for direct comparison with \dl\ and XMASS. 
The dashed line corresponds to the case of zero modulation amplitudes for comparison with data. }
\label{fig:highe}
\end{figure}

Using the same analysis method, 
the annual modulation study was extended up to 26~\keVee. 
Above this energy window the ER spectrum begins to be contaminated 
by xenon X-rays and residual \krthree\ decays. 
In addition, the SS cut efficiency was observed to drop slightly from 100\% above energies of 15-20~\keVee, 
as a result of background electron pulses following primary S2s being tagged as additional S2s. 
However, the SS efficiency below 26~\keVee\ was evaluated to be above 95\%, 
and the time dependence is less than 5\%. 
Figure~\ref{fig:highe} (top) shows the 90\% CL contour for each data group in the modulation parameter space. 
For all the data divided in 4~\keVee\ energy bins, 
no annual modulation amplitude above 1.1~cpd/\keVee/tonne or above 2 sigma deviation from zero is observed. 
For easy comparison with other experiments, 
Figure~\ref{fig:highe} (bottom) shows the best-fit modulation amplitudes as a function of energy, 
with the modulation phase fixed at 152 days. 

\subsection{Diurnal modulation}
\label{subsec:dimod}

Due to the small amplitude, 
diurnal modulations have not been widely discussed in dark matter experiments. 
To date, the only experimental search for diurnal modulations was from the \dl\ experiment, 
which concluded that the signal was too small to be observed~\cite{DAMA2014_DiurnalMod}. 
The LUX experiment has achieved a total event rate $\sim$500 times lower than that of \dl, 
which may enable a sensitive search for diurnal modulations. 
Particularly, dissipative dark matter models including self interactions 
typically predict a larger effect from the Earth's spin~\cite{Clarke2016_PlasmaDM, Foot2015_DiurnalMod}. 
In such models, a significant amount of dark matter particles may be captured by the Earth 
due to dark matter-matter scattering and also the self-interaction of dark matter. 
The amount of captured dark matter within the Earth may maintain a dynamic equilibrium 
between the loss of previously captured dark matter to the halo wind 
and newly captured dark matter. 
This exchange of dark matter content may occur close to the surface of the Earth, 
and therefore lead to relatively large diurnal modulation amplitudes in dark matter direct detection experiments as the Earth spins. 
This section discusses such a search for diurnal modulations using the same 2-6~\keVee\ LUX data set 
used in the annual modulation analysis discussed above. 

\begin{figure}[h!]
\centering
\includegraphics[width=.45\textwidth]{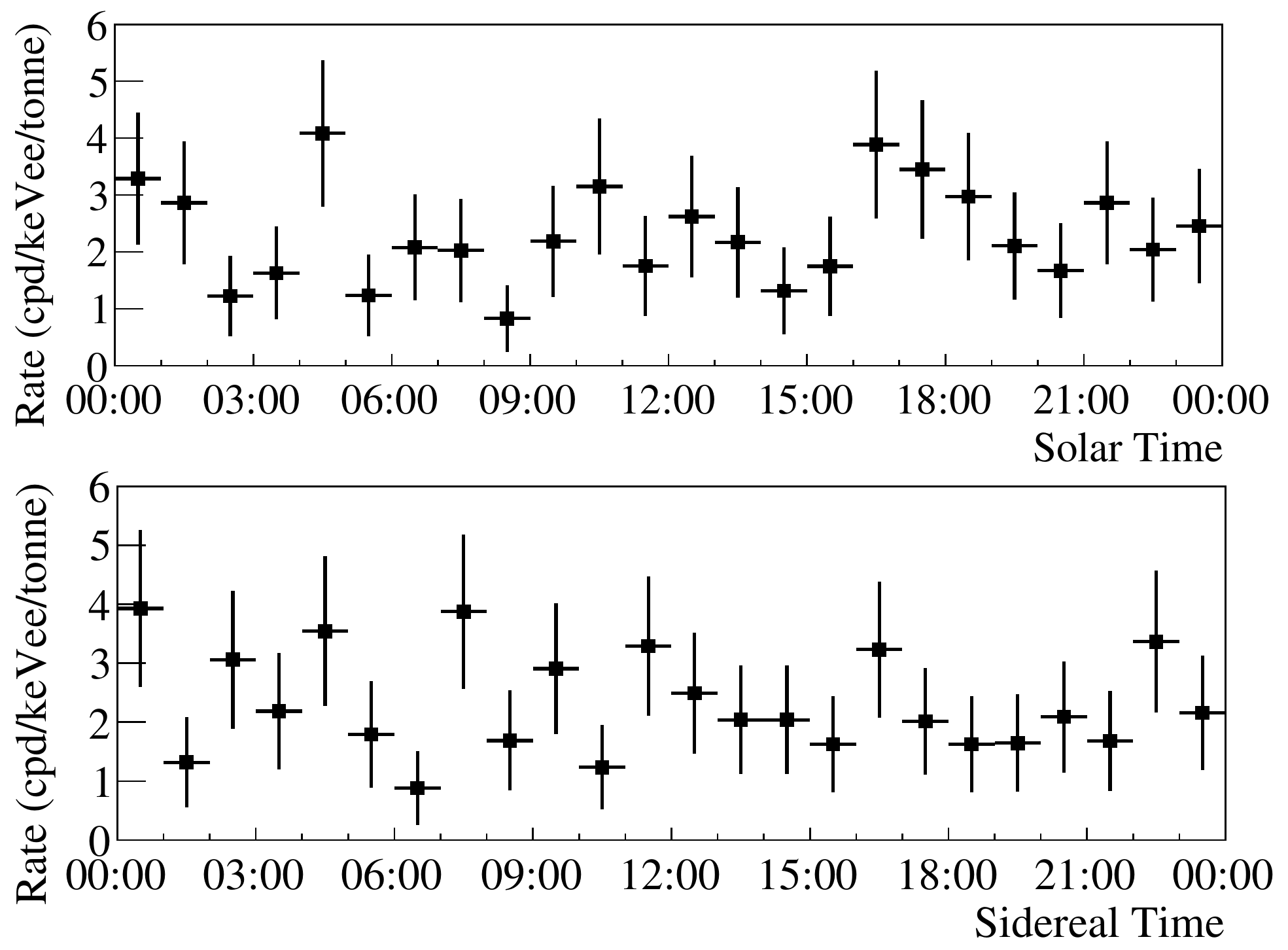}
\caption{The observed ER event rate in the LUX detector as a function of time of the day (presented in hours); 
the rates were calculated for both solar time, i.e., Mountain time (top), and local sidereal time (bottom). }
\label{fig:dmod}
\end{figure}

Figure~\ref{fig:dmod} shows the observed ER event rate between 2 and 6~\keVee\ at different times of the day, 
calculated with respect to both solar time (top) and sidereal time at 103.77$^{\circ}$ W (bottom). 
Because experimental conditions may vary with time in the solar day and could impact the background levels, 
a solar time analysis can help verify the diurnal stability of the experiment. 
No significant time dependence of the event rate is observed in either group of data.  
We calculated the average event rates during the day (night) to be 2.26 (2.37)~cpd/\keVee/tonne 
and 2.28 (2.36)~cpd/\keVee/tonne, for solar time and sidereal time, respectively. 
Similarly, the average rate in the morning (evening) were calculated to be 2.19 (2.44)~cpd/\keVee/tonne 
and 2.48 (2.16)~cpd/\keVee/tonne, for solar time and sidereal time, respectively. 
For convenience, we defined day, night, morning and evening in local sidereal time in analogy to 
that of local solar time using the 24 hours convention. 
As suggested by \cite{Clarke2016_PlasmaDM} and  \cite{Foot2015_DiurnalMod}, 
the dark matter interaction rate in certain dark matter models may exhibit a significant difference  
in the sidereal day/night or morning/evening. 

Due to the lack of a generic diurnal modulation model, 
a full modulation analysis as that in the annual modulation search was not carried out. 
Instead, a simple 12-hour asymmetry factor
$$ \mathcal{A}_{t} = \frac{R_t-\bar{R_t}}{R_t+\bar{R_t}}$$
was calculated, 
where $R_t$ is the average event rate in a 12 hour time window centered at the time of interest, 
and $\bar{R_t}$ is the average rate in the supplemental 12 hour window. 
For example, $\mathcal{A}_{12}$, or the asymmetry factor at noon, 
would represent a day-night asymmetry in the event rate. 
A value of $\mathcal{A}_{12}>0$ would indicate a higher event rate during the day, 
and $\mathcal{A}_{12}<0$ indicates the opposite. 
In addition, a non-zero $\mathcal{A}_{12}$ value in solar time would suggest the existence of a time-dependent background in the experiment 
and the sensitivity of a dark matter diurnal search may be compromised. 

For the situation of solar time, 
the day-night asymmetry is calculated to be -5.3$\pm$8.7\%, 
and the morning-evening asymmetry is calculated to be -2.5$\pm$8.7\%; 
both are consistent with zero. 
For sidereal time, 
the day-night asymmetry is calculated to be -1.7$\pm$8.7\%, 
and the morning-evening asymmetry is calculated to be 6.7$\pm$8.8\%. 
The uncertainties in both results represent the statistic uncertainties only. 
In conclusion, at the sensitivity level of $\sim$9\% or $\sim$0.2~cpd/\keVee/tonne, 
we report no observation of statistically significant diurnal modulation features in the LUX data, 
either in solar time or in sidereal time. 
Due to the limited sensitivity from low statistics and the lack of generic modulation predictions, 
the diurnal modulation search was not extended to other energy regions. 

\section{Conclusion}
\label{sec:concl}

We carried out a search for annual and diurnal rate modulations in the ER events  
collected with the LUX dark matter detector between 2013 and 2016. 
Despite a significant time dependence in the experimental operation conditions, 
we achieved a low and stable event rate for this analysis by developing robust cuts and corrections. 
We report no significant annual modulation signatures in the energy window of 2-26~\keVee\ in the LUX data. 
This LUX result consisted of the most stringent annual modulation search between 2 and 6~\keVee\ 
by demonstrating the lowest 90\% CL limits in modulation amplitude, 
and the best fit modulation parameter is in 9.2$\sigma$ tension with that reported by \dl. 
For the diurnal modulation search, 
this analysis disfavors any day-night asymmetry or morning-evening asymmetry above 0.2~cpd/\keVee/tonne level. 

\begin{acknowledgments}

 This work was partially supported by the U.S. Department of Energy (DOE) under award numbers DE-FG02-08ER41549, DE-FG02-91ER40688, DE-FG02-95ER40917, DE-FG02-91ER40674, DE-NA0000979, DE-FG02-11ER41738, DE-SC0006605, DE-AC02-05CH11231, DE-AC52-07NA27344, DE-FG01-91ER40618 and DE-SC0010010; the U.S. National Science Foundation under award numbers PHYS-0750671, PHY-0801536, PHY-1004661, PHY-1102470, PHY-1003660, PHY-1312561, PHY-1347449, PHY-1505868, PHY-1636738, PHY-0919261; the Research Corporation grant RA0350; the Center for Ultra-low Background Experiments in the Dakotas (CUBED); and the South Dakota School of Mines and Technology (SDSMT). LIP-Coimbra acknowledges funding from Funda\c c\~ao para a Ci\^encia e Tecnologia(FCT) through the project-grant PTDC/FIS-NUC/1525/2014. Imperial College and Brown University thank the UK Royal Society for travel funds under the International Exchange Scheme (IE120804). The UK groups acknowledge institutional support from Imperial College London, University College London and Edinburgh University, and from the Science \& Technology Facilities Council for PhD studentship ST/K502042/1 (AB). The University of Edinburgh is a charitable body, registered in Scotland, with registration number SC005336.

This research was conducted using computational resources and services at the Center for Computation and Visualization, Brown University.

The authors acknowledge the work of the following engineers who played important roles during the design, construction, commissioning, and operation phases of LUX: S. Dardin from Berkeley, B. Holbrook, R. Gerhard, and J. Thomson from UC Davis, and G. Mok, J. Bauer, and D. Carr from Livermore.

The authors gratefully acknowledge the logistical and technical support and the access to laboratory infrastructure provided to us by the Sanford Underground Research Facility (SURF) and its personnel at Lead, South Dakota. SURF was developed by the South Dakota Science and Technology authority, with an important philanthropic donation from T. Denny Sanford, and is operated by Lawrence Berkeley National Laboratory for the Department of Energy, Office of High Energy Physics. 

LLNL is operated by Lawrence Livermore National Security, LLC, for the U.S. Department of Energy, National Nuclear Security Administration under Contract DE-AC52-07NA27344.
%The LLNL release number for this manuscript is LLNL-JRNL-757487. 
\end{acknowledgments}

\bibliographystyle{apsrev}
\bibliography{biblio}

\begin{thebibliography}{42}
\expandafter\ifx\csname natexlab\endcsname\relax\def\natexlab#1{#1}\fi
\expandafter\ifx\csname bibnamefont\endcsname\relax
  \def\bibnamefont#1{#1}\fi
\expandafter\ifx\csname bibfnamefont\endcsname\relax
  \def\bibfnamefont#1{#1}\fi
\expandafter\ifx\csname citenamefont\endcsname\relax
  \def\citenamefont#1{#1}\fi
\expandafter\ifx\csname url\endcsname\relax
  \def\url#1{\texttt{#1}}\fi
\expandafter\ifx\csname urlprefix\endcsname\relax\def\urlprefix{URL }\fi
\providecommand{\bibinfo}[2]{#2}
\providecommand{\eprint}[2][]{\url{#2}}

\bibitem[{\citenamefont{Akerib et~al.}(2017{\natexlab{a}})\citenamefont{Akerib,
  Alsum, Ara\'ujo, Bai, Bailey, Balajthy, Beltrame, Bernard, Bernstein,
  Biesiadzinski et~al.}}]{LUX2016_Run3_4}
\bibinfo{author}{\bibfnamefont{D.~S.} \bibnamefont{Akerib}},
  \bibinfo{author}{\bibfnamefont{S.}~\bibnamefont{Alsum}},
  \bibinfo{author}{\bibfnamefont{H.~M.} \bibnamefont{Ara\'ujo}},
  \bibinfo{author}{\bibfnamefont{X.}~\bibnamefont{Bai}},
  \bibinfo{author}{\bibfnamefont{A.~J.} \bibnamefont{Bailey}},
  \bibinfo{author}{\bibfnamefont{J.}~\bibnamefont{Balajthy}},
  \bibinfo{author}{\bibfnamefont{P.}~\bibnamefont{Beltrame}},
  \bibinfo{author}{\bibfnamefont{E.~P.} \bibnamefont{Bernard}},
  \bibinfo{author}{\bibfnamefont{A.}~\bibnamefont{Bernstein}},
  \bibinfo{author}{\bibfnamefont{T.~P.} \bibnamefont{Biesiadzinski}},
  \bibnamefont{et~al.} (\bibinfo{collaboration}{LUX Collaboration}),
  \href{http://dx.doi.org/10.1103/PhysRevLett.118.021303}{\bibinfo{journal}{Phys.
  Rev. Lett.}, \textbf{\bibinfo{volume}{118}},
  \bibinfo{pages}{021303}\bibinfo{year}{
  (\bibinfo{year}{2017}{\natexlab{a}})}}.

\bibitem[{\citenamefont{Aprile et~al.}(2017{\natexlab{a}})\citenamefont{Aprile,
  Aalbers, Agostini, Alfonsi, Amaro, Anthony, Arneodo, Barrow, Baudis,
  Bauermeister et~al.}}]{XENON1T_2017}
\bibinfo{author}{\bibfnamefont{E.}~\bibnamefont{Aprile}},
  \bibinfo{author}{\bibfnamefont{J.}~\bibnamefont{Aalbers}},
  \bibinfo{author}{\bibfnamefont{F.}~\bibnamefont{Agostini}},
  \bibinfo{author}{\bibfnamefont{M.}~\bibnamefont{Alfonsi}},
  \bibinfo{author}{\bibfnamefont{F.~D.} \bibnamefont{Amaro}},
  \bibinfo{author}{\bibfnamefont{M.}~\bibnamefont{Anthony}},
  \bibinfo{author}{\bibfnamefont{F.}~\bibnamefont{Arneodo}},
  \bibinfo{author}{\bibfnamefont{P.}~\bibnamefont{Barrow}},
  \bibinfo{author}{\bibfnamefont{L.}~\bibnamefont{Baudis}},
  \bibinfo{author}{\bibfnamefont{B.}~\bibnamefont{Bauermeister}},
  \bibnamefont{et~al.} (\bibinfo{collaboration}{XENON Collaboration}),
  \href{http://dx.doi.org/10.1103/PhysRevLett.119.181301}{\bibinfo{journal}{Phys.
  Rev. Lett.}, \textbf{\bibinfo{volume}{119}},
  \bibinfo{pages}{181301}\bibinfo{year}{
  (\bibinfo{year}{2017}{\natexlab{a}})}}.

\bibitem[{\citenamefont{Cui et~al.}(2017)\citenamefont{Cui, Abdukerim, Chen,
  Chen, Chen, Dong, Fang, Fu, Giboni, Giuliani et~al.}}]{PandaX2017_DM}
\bibinfo{author}{\bibfnamefont{X.}~\bibnamefont{Cui}},
  \bibinfo{author}{\bibfnamefont{A.}~\bibnamefont{Abdukerim}},
  \bibinfo{author}{\bibfnamefont{W.}~\bibnamefont{Chen}},
  \bibinfo{author}{\bibfnamefont{X.}~\bibnamefont{Chen}},
  \bibinfo{author}{\bibfnamefont{Y.}~\bibnamefont{Chen}},
  \bibinfo{author}{\bibfnamefont{B.}~\bibnamefont{Dong}},
  \bibinfo{author}{\bibfnamefont{D.}~\bibnamefont{Fang}},
  \bibinfo{author}{\bibfnamefont{C.}~\bibnamefont{Fu}},
  \bibinfo{author}{\bibfnamefont{K.}~\bibnamefont{Giboni}},
  \bibinfo{author}{\bibfnamefont{F.}~\bibnamefont{Giuliani}},
  \bibnamefont{et~al.} (\bibinfo{collaboration}{PandaX-II Collaboration}),
  \href{http://dx.doi.org/10.1103/PhysRevLett.119.181302}{\bibinfo{journal}{Phys.
  Rev. Lett.}, \textbf{\bibinfo{volume}{119}},
  \bibinfo{pages}{181302}\bibinfo{year}{ (\bibinfo{year}{2017})}}.

\bibitem[{\citenamefont{Agnese et~al.}(2018{\natexlab{a}})\citenamefont{Agnese,
  Aramaki, Arnquist, Baker, Balakishiyeva, Banik, Barker, Basu~Thakur, Bauer,
  Binder et~al.}}]{CDMS2018_DM}
\bibinfo{author}{\bibfnamefont{R.}~\bibnamefont{Agnese}},
  \bibinfo{author}{\bibfnamefont{T.}~\bibnamefont{Aramaki}},
  \bibinfo{author}{\bibfnamefont{I.~J.} \bibnamefont{Arnquist}},
  \bibinfo{author}{\bibfnamefont{W.}~\bibnamefont{Baker}},
  \bibinfo{author}{\bibfnamefont{D.}~\bibnamefont{Balakishiyeva}},
  \bibinfo{author}{\bibfnamefont{S.}~\bibnamefont{Banik}},
  \bibinfo{author}{\bibfnamefont{D.}~\bibnamefont{Barker}},
  \bibinfo{author}{\bibfnamefont{R.}~\bibnamefont{Basu~Thakur}},
  \bibinfo{author}{\bibfnamefont{D.~A.} \bibnamefont{Bauer}},
  \bibinfo{author}{\bibfnamefont{T.}~\bibnamefont{Binder}},
  \bibnamefont{et~al.} (\bibinfo{collaboration}{SuperCDMS Collaboration}),
  \href{http://dx.doi.org/10.1103/PhysRevLett.120.061802}{\bibinfo{journal}{Phys.
  Rev. Lett.}, \textbf{\bibinfo{volume}{120}},
  \bibinfo{pages}{061802}\bibinfo{year}{
  (\bibinfo{year}{2018}{\natexlab{a}})}}.

\bibitem[{\citenamefont{Kopp et~al.}(2009)\citenamefont{Kopp, Niro, Schwetz,
  and Zupan}}]{Kopp2009_LeptDM}
\bibinfo{author}{\bibfnamefont{J.}~\bibnamefont{Kopp}},
  \bibinfo{author}{\bibfnamefont{V.}~\bibnamefont{Niro}},
  \bibinfo{author}{\bibfnamefont{T.}~\bibnamefont{Schwetz}}, \bibnamefont{and}
  \bibinfo{author}{\bibfnamefont{J.}~\bibnamefont{Zupan}},
  \href{http://dx.doi.org/10.1103/PhysRevD.80.083502}{\bibinfo{journal}{Phys.
  Rev. D}, \textbf{\bibinfo{volume}{80}},
  \bibinfo{pages}{083502}\bibinfo{year}{ (\bibinfo{year}{2009})}}.

\bibitem[{\citenamefont{Fox and Poppitz}(2009)}]{Fox2009_LeptDM}
\bibinfo{author}{\bibfnamefont{P.~J.} \bibnamefont{Fox}} \bibnamefont{and}
  \bibinfo{author}{\bibfnamefont{E.}~\bibnamefont{Poppitz}},
  \href{http://dx.doi.org/10.1103/PhysRevD.79.083528}{\bibinfo{journal}{Phys.
  Rev. D}, \textbf{\bibinfo{volume}{79}},
  \bibinfo{pages}{083528}\bibinfo{year}{ (\bibinfo{year}{2009})}}.

\bibitem[{\citenamefont{Drukier et~al.}(1986)\citenamefont{Drukier, Freese, and
  Spergel}}]{Drukier1986_Modulation}
\bibinfo{author}{\bibfnamefont{A.~K.} \bibnamefont{Drukier}},
  \bibinfo{author}{\bibfnamefont{K.}~\bibnamefont{Freese}}, \bibnamefont{and}
  \bibinfo{author}{\bibfnamefont{D.~N.} \bibnamefont{Spergel}},
  \href{http://dx.doi.org/10.1103/PhysRevD.33.3495}{\bibinfo{journal}{Phys.
  Rev. D}, \textbf{\bibinfo{volume}{33}}, \bibinfo{pages}{3495}\bibinfo{year}{
  (\bibinfo{year}{1986})}}.

\bibitem[{\citenamefont{Lewin and Smith}(1996)}]{Lewin_review_1996}
\bibinfo{author}{\bibfnamefont{J.~D.} \bibnamefont{Lewin}} \bibnamefont{and}
  \bibinfo{author}{\bibfnamefont{P.~F.} \bibnamefont{Smith}},
  \href{http://dx.doi.org/10.1016/S0927-6505(96)00047-3}{\bibinfo{journal}{Astroparticle
  Physics}, \textbf{\bibinfo{volume}{6}}, \bibinfo{pages}{87}\bibinfo{year}{
  (\bibinfo{year}{1996})}}, ISSN \bibinfo{issn}{0927-6505}.

\bibitem[{\citenamefont{Freese et~al.}(2013)\citenamefont{Freese, Lisanti, and
  Savage}}]{Freese2013_Modulation}
\bibinfo{author}{\bibfnamefont{K.}~\bibnamefont{Freese}},
  \bibinfo{author}{\bibfnamefont{M.}~\bibnamefont{Lisanti}}, \bibnamefont{and}
  \bibinfo{author}{\bibfnamefont{C.}~\bibnamefont{Savage}},
  \href{http://dx.doi.org/10.1103/RevModPhys.85.1561}{\bibinfo{journal}{Rev.
  Mod. Phys.}, \textbf{\bibinfo{volume}{85}},
  \bibinfo{pages}{1561}\bibinfo{year}{ (\bibinfo{year}{2013})}}.

\bibitem[{\citenamefont{Semertzidis and
  Vergados}(2015)}]{Semertzidis2015_AxionDMModulation}
\bibinfo{author}{\bibfnamefont{Y.}~\bibnamefont{Semertzidis}} \bibnamefont{and}
  \bibinfo{author}{\bibfnamefont{J.}~\bibnamefont{Vergados}},
  \href{http://dx.doi.org/https://doi.org/10.1016/j.nuclphysb.2015.06.014}{\bibinfo{journal}{Nuclear
  Physics B}, \textbf{\bibinfo{volume}{897}}, \bibinfo{pages}{821
  }\bibinfo{year}{ (\bibinfo{year}{2015})}}, ISSN \bibinfo{issn}{0550-3213}.

\bibitem[{\citenamefont{Foot}(2004)}]{Foot2004_DAMAMirrorDM}
\bibinfo{author}{\bibfnamefont{R.}~\bibnamefont{Foot}},
  \href{http://dx.doi.org/10.1103/PhysRevD.69.036001}{\bibinfo{journal}{Phys.
  Rev. D}, \textbf{\bibinfo{volume}{69}},
  \bibinfo{pages}{036001}\bibinfo{year}{ (\bibinfo{year}{2004})}}.

\bibitem[{\citenamefont{Foot}(2013)}]{Foot2013_MirrorDM}
\bibinfo{author}{\bibfnamefont{R.}~\bibnamefont{Foot}},
  \href{http://dx.doi.org/10.1103/PhysRevD.88.023520}{\bibinfo{journal}{Phys.
  Rev. D}, \textbf{\bibinfo{volume}{88}},
  \bibinfo{pages}{023520}\bibinfo{year}{ (\bibinfo{year}{2013})}}.

\bibitem[{\citenamefont{Clarke and Foot}(2016)}]{Clarke2016_PlasmaDM}
\bibinfo{author}{\bibfnamefont{J.}~\bibnamefont{Clarke}} \bibnamefont{and}
  \bibinfo{author}{\bibfnamefont{R.}~\bibnamefont{Foot}},
  \href{http://stacks.iop.org/1475-7516/2016/i=01/a=029}{\bibinfo{journal}{Journal
  of Cosmology and Astroparticle Physics}, \textbf{\bibinfo{volume}{2016}},
  \bibinfo{pages}{029}\bibinfo{year}{ (\bibinfo{year}{2016})}}.

\bibitem[{\citenamefont{Bernabei et~al.}(2013)\citenamefont{Bernabei, Belli,
  Cappella, Caracciolo, Castellano, Cerulli, Dai, d'Angelo, d'Angelo, Marco
  et~al.}}]{DAMA2013_Phase1}
\bibinfo{author}{\bibfnamefont{R.}~\bibnamefont{Bernabei}},
  \bibinfo{author}{\bibfnamefont{P.}~\bibnamefont{Belli}},
  \bibinfo{author}{\bibfnamefont{F.}~\bibnamefont{Cappella}},
  \bibinfo{author}{\bibfnamefont{V.}~\bibnamefont{Caracciolo}},
  \bibinfo{author}{\bibfnamefont{S.}~\bibnamefont{Castellano}},
  \bibinfo{author}{\bibfnamefont{R.}~\bibnamefont{Cerulli}},
  \bibinfo{author}{\bibfnamefont{C.~J.} \bibnamefont{Dai}},
  \bibinfo{author}{\bibfnamefont{A.}~\bibnamefont{d'Angelo}},
  \bibinfo{author}{\bibfnamefont{S.}~\bibnamefont{d'Angelo}},
  \bibinfo{author}{\bibfnamefont{A.~D.} \bibnamefont{Marco}},
  \bibnamefont{et~al.},
  \href{http://dx.doi.org/10.1140/epjc/s10052-013-2648-7}{\bibinfo{journal}{The
  European Physical Journal C}, \textbf{\bibinfo{volume}{73}},
  \bibinfo{pages}{1}\bibinfo{year}{ (\bibinfo{year}{2013})}}, ISSN
  \bibinfo{issn}{1434-6044, 1434-6052}.

\bibitem[{\citenamefont{\relax
  DAMA/LIBRA~Collaboration}(2018)}]{DAMA2018_Phase2}
\bibinfo{author}{\bibnamefont{\relax DAMA/LIBRA~Collaboration}},
  \href{http://arxiv.org/abs/arXiv:1805.10486}{\bibinfo{journal}{arXiv}:arXiv:1805.10486\bibinfo{year}{
  (\bibinfo{year}{2018})}}.

\bibitem[{\citenamefont{Aalseth et~al.}(2011)\citenamefont{Aalseth, Barbeau,
  Colaresi, Collar, Diaz~Leon, Fast, Fields, Hossbach, Keillor, Kephart
  et~al.}}]{CoGeNT2011_Mod}
\bibinfo{author}{\bibfnamefont{C.~E.} \bibnamefont{Aalseth}},
  \bibinfo{author}{\bibfnamefont{P.~S.} \bibnamefont{Barbeau}},
  \bibinfo{author}{\bibfnamefont{J.}~\bibnamefont{Colaresi}},
  \bibinfo{author}{\bibfnamefont{J.~I.} \bibnamefont{Collar}},
  \bibinfo{author}{\bibfnamefont{J.}~\bibnamefont{Diaz~Leon}},
  \bibinfo{author}{\bibfnamefont{J.~E.} \bibnamefont{Fast}},
  \bibinfo{author}{\bibfnamefont{N.}~\bibnamefont{Fields}},
  \bibinfo{author}{\bibfnamefont{T.~W.} \bibnamefont{Hossbach}},
  \bibinfo{author}{\bibfnamefont{M.~E.} \bibnamefont{Keillor}},
  \bibinfo{author}{\bibfnamefont{J.~D.} \bibnamefont{Kephart}},
  \bibnamefont{et~al.} (\bibinfo{collaboration}{CoGeNT Collaboration}),
  \href{http://dx.doi.org/10.1103/PhysRevLett.107.141301}{\bibinfo{journal}{Phys.
  Rev. Lett.}, \textbf{\bibinfo{volume}{107}},
  \bibinfo{pages}{141301}\bibinfo{year}{ (\bibinfo{year}{2011})}}.

\bibitem[{\citenamefont{Agnese et~al.}(2018{\natexlab{b}})\citenamefont{Agnese,
  Anderson, Aralis, Aramaki, Arnquist, Baker, Balakishiyeva, Barker,
  Basu~Thakur, Bauer et~al.}}]{CDMS2018_LowMass}
\bibinfo{author}{\bibfnamefont{R.}~\bibnamefont{Agnese}},
  \bibinfo{author}{\bibfnamefont{A.~J.} \bibnamefont{Anderson}},
  \bibinfo{author}{\bibfnamefont{T.}~\bibnamefont{Aralis}},
  \bibinfo{author}{\bibfnamefont{T.}~\bibnamefont{Aramaki}},
  \bibinfo{author}{\bibfnamefont{I.~J.} \bibnamefont{Arnquist}},
  \bibinfo{author}{\bibfnamefont{W.}~\bibnamefont{Baker}},
  \bibinfo{author}{\bibfnamefont{D.}~\bibnamefont{Balakishiyeva}},
  \bibinfo{author}{\bibfnamefont{D.}~\bibnamefont{Barker}},
  \bibinfo{author}{\bibfnamefont{R.}~\bibnamefont{Basu~Thakur}},
  \bibinfo{author}{\bibfnamefont{D.~A.} \bibnamefont{Bauer}},
  \bibnamefont{et~al.} (\bibinfo{collaboration}{SuperCDMS Collaboration}),
  \href{http://dx.doi.org/10.1103/PhysRevD.97.022002}{\bibinfo{journal}{Phys.
  Rev. D}, \textbf{\bibinfo{volume}{97}},
  \bibinfo{pages}{022002}\bibinfo{year}{
  (\bibinfo{year}{2018}{\natexlab{b}})}}.

\bibitem[{\citenamefont{Angloher et~al.}(2012)\citenamefont{Angloher, Bauer,
  Bavykina, Bento, Bucci, Ciemniak, Deuter, von Feilitzsch, Hauff, Huff
  et~al.}}]{CRESST2012_Excess}
\bibinfo{author}{\bibfnamefont{G.}~\bibnamefont{Angloher}},
  \bibinfo{author}{\bibfnamefont{M.}~\bibnamefont{Bauer}},
  \bibinfo{author}{\bibfnamefont{I.}~\bibnamefont{Bavykina}},
  \bibinfo{author}{\bibfnamefont{A.}~\bibnamefont{Bento}},
  \bibinfo{author}{\bibfnamefont{C.}~\bibnamefont{Bucci}},
  \bibinfo{author}{\bibfnamefont{C.}~\bibnamefont{Ciemniak}},
  \bibinfo{author}{\bibfnamefont{G.}~\bibnamefont{Deuter}},
  \bibinfo{author}{\bibfnamefont{F.}~\bibnamefont{von Feilitzsch}},
  \bibinfo{author}{\bibfnamefont{D.}~\bibnamefont{Hauff}},
  \bibinfo{author}{\bibfnamefont{P.}~\bibnamefont{Huff}}, \bibnamefont{et~al.},
   \href{http://dx.doi.org/10.1140/epjc/s10052-012-1971-8}{\bibinfo{journal}{The
  European Physical Journal C}, \textbf{\bibinfo{volume}{72}},
  \bibinfo{pages}{1971}\bibinfo{year}{ (\bibinfo{year}{2012})}}, ISSN
  \bibinfo{issn}{1434-6052}.

\bibitem[{\citenamefont{Bernabei et~al.}(2008)\citenamefont{Bernabei, Belli,
  Cappella, Cerulli, Dai, d'Angelo, He, Incicchitti, Kuang, Ma
  et~al.}}]{DAMA2008_CombinedAnalysis}
\bibinfo{author}{\bibfnamefont{R.}~\bibnamefont{Bernabei}},
  \bibinfo{author}{\bibfnamefont{P.}~\bibnamefont{Belli}},
  \bibinfo{author}{\bibfnamefont{F.}~\bibnamefont{Cappella}},
  \bibinfo{author}{\bibfnamefont{R.}~\bibnamefont{Cerulli}},
  \bibinfo{author}{\bibfnamefont{C.~J.} \bibnamefont{Dai}},
  \bibinfo{author}{\bibfnamefont{A.}~\bibnamefont{d'Angelo}},
  \bibinfo{author}{\bibfnamefont{H.~L.} \bibnamefont{He}},
  \bibinfo{author}{\bibfnamefont{A.}~\bibnamefont{Incicchitti}},
  \bibinfo{author}{\bibfnamefont{H.~H.} \bibnamefont{Kuang}},
  \bibinfo{author}{\bibfnamefont{J.~M.} \bibnamefont{Ma}},
  \bibnamefont{et~al.},
  \href{http://dx.doi.org/10.1140/epjc/s10052-008-0662-y}{\bibinfo{journal}{The
  European Physical Journal C}, \textbf{\bibinfo{volume}{56}},
  \bibinfo{pages}{333}\bibinfo{year}{ (\bibinfo{year}{2008})}}, ISSN
  \bibinfo{issn}{1434-6052}.

\bibitem[{\citenamefont{Aprile et~al.}(2018)\citenamefont{Aprile, Aalbers,
  Agostini, Alfonsi, Althueser, Amaro, Anthony, Arneodo, Baudis, Bauermeister
  et~al.}}]{XENON1T_2018}
\bibinfo{author}{\bibfnamefont{E.}~\bibnamefont{Aprile}},
  \bibinfo{author}{\bibfnamefont{J.}~\bibnamefont{Aalbers}},
  \bibinfo{author}{\bibfnamefont{F.}~\bibnamefont{Agostini}},
  \bibinfo{author}{\bibfnamefont{M.}~\bibnamefont{Alfonsi}},
  \bibinfo{author}{\bibfnamefont{L.}~\bibnamefont{Althueser}},
  \bibinfo{author}{\bibfnamefont{F.~D.} \bibnamefont{Amaro}},
  \bibinfo{author}{\bibfnamefont{M.}~\bibnamefont{Anthony}},
  \bibinfo{author}{\bibfnamefont{F.}~\bibnamefont{Arneodo}},
  \bibinfo{author}{\bibfnamefont{L.}~\bibnamefont{Baudis}},
  \bibinfo{author}{\bibfnamefont{B.}~\bibnamefont{Bauermeister}},
  \bibnamefont{et~al.} (\bibinfo{collaboration}{XENON Collaboration 7}),
  \href{http://dx.doi.org/10.1103/PhysRevLett.121.111302}{\bibinfo{journal}{Phys.
  Rev. Lett.}, \textbf{\bibinfo{volume}{121}},
  \bibinfo{pages}{111302}\bibinfo{year}{ (\bibinfo{year}{2018})}}.

\bibitem[{\citenamefont{{\relax The XENON
  Collaboration}}(2015)}]{XENON2015_Leptophilic}
\bibinfo{author}{\bibnamefont{{\relax The XENON Collaboration}}},
  \href{http://dx.doi.org/10.1126/science.aab2069}{\bibinfo{journal}{Science},
  \textbf{\bibinfo{volume}{349}}, \bibinfo{pages}{851}\bibinfo{year}{
  (\bibinfo{year}{2015})}}, ISSN \bibinfo{issn}{0036-8075}.

\bibitem[{\citenamefont{Bernabei et~al.}(2014)\citenamefont{Bernabei, Belli,
  Cappella, Caracciolo, Castellano, Cerulli, Dai, d'Angelo, d'Angelo, Di~Marco
  et~al.}}]{DAMA2014_DiurnalMod}
\bibinfo{author}{\bibfnamefont{R.}~\bibnamefont{Bernabei}},
  \bibinfo{author}{\bibfnamefont{P.}~\bibnamefont{Belli}},
  \bibinfo{author}{\bibfnamefont{F.}~\bibnamefont{Cappella}},
  \bibinfo{author}{\bibfnamefont{V.}~\bibnamefont{Caracciolo}},
  \bibinfo{author}{\bibfnamefont{S.}~\bibnamefont{Castellano}},
  \bibinfo{author}{\bibfnamefont{R.}~\bibnamefont{Cerulli}},
  \bibinfo{author}{\bibfnamefont{C.~J.} \bibnamefont{Dai}},
  \bibinfo{author}{\bibfnamefont{A.}~\bibnamefont{d'Angelo}},
  \bibinfo{author}{\bibfnamefont{S.}~\bibnamefont{d'Angelo}},
  \bibinfo{author}{\bibfnamefont{A.}~\bibnamefont{Di~Marco}},
  \bibnamefont{et~al.},
  \href{http://dx.doi.org/10.1140/epjc/s10052-014-2827-1}{\bibinfo{journal}{The
  European Physical Journal C}, \textbf{\bibinfo{volume}{74}},
  \bibinfo{pages}{2827}\bibinfo{year}{ (\bibinfo{year}{2014})}}, ISSN
  \bibinfo{issn}{1434-6052}.

\bibitem[{\citenamefont{Foot and Vagnozzi}(2015)}]{Foot2015_DiurnalMod}
\bibinfo{author}{\bibfnamefont{R.}~\bibnamefont{Foot}} \bibnamefont{and}
  \bibinfo{author}{\bibfnamefont{S.}~\bibnamefont{Vagnozzi}},
  \href{http://dx.doi.org/https://doi.org/10.1016/j.physletb.2015.06.063}{\bibinfo{journal}{Physics
  Letters B}, \textbf{\bibinfo{volume}{748}}, \bibinfo{pages}{61
  }\bibinfo{year}{ (\bibinfo{year}{2015})}}, ISSN \bibinfo{issn}{0370-2693}.

\bibitem[{\citenamefont{Akerib et~al.}(2013)\citenamefont{Akerib, Bai,
  Bedikian, Bernard, Bernstein, Bolozdynya, Bradley, Byram, Cahn, Camp
  et~al.}}]{LUX2012_Detector}
\bibinfo{author}{\bibfnamefont{D.}~\bibnamefont{Akerib}},
  \bibinfo{author}{\bibfnamefont{X.}~\bibnamefont{Bai}},
  \bibinfo{author}{\bibfnamefont{S.}~\bibnamefont{Bedikian}},
  \bibinfo{author}{\bibfnamefont{E.}~\bibnamefont{Bernard}},
  \bibinfo{author}{\bibfnamefont{A.}~\bibnamefont{Bernstein}},
  \bibinfo{author}{\bibfnamefont{A.}~\bibnamefont{Bolozdynya}},
  \bibinfo{author}{\bibfnamefont{A.}~\bibnamefont{Bradley}},
  \bibinfo{author}{\bibfnamefont{D.}~\bibnamefont{Byram}},
  \bibinfo{author}{\bibfnamefont{S.}~\bibnamefont{Cahn}},
  \bibinfo{author}{\bibfnamefont{C.}~\bibnamefont{Camp}}, \bibnamefont{et~al.},
   \href{http://dx.doi.org/https://doi.org/10.1016/j.nima.2012.11.135}{\bibinfo{journal}{Nuclear
  Instruments and Methods in Physics Research Section A: Accelerators,
  Spectrometers, Detectors and Associated Equipment},
  \textbf{\bibinfo{volume}{704}}, \bibinfo{pages}{111 }\bibinfo{year}{
  (\bibinfo{year}{2013})}}, ISSN \bibinfo{issn}{0168-9002}.

\bibitem[{\citenamefont{Akerib et~al.}(2016{\natexlab{a}})\citenamefont{Akerib,
  Ara\'ujo, Bai, Bailey, Balajthy, Beltrame, Bernard, Bernstein, Biesiadzinski,
  Boulton et~al.}}]{LUX2015_Reanalysis}
\bibinfo{author}{\bibfnamefont{D.~S.} \bibnamefont{Akerib}},
  \bibinfo{author}{\bibfnamefont{H.~M.} \bibnamefont{Ara\'ujo}},
  \bibinfo{author}{\bibfnamefont{X.}~\bibnamefont{Bai}},
  \bibinfo{author}{\bibfnamefont{A.~J.} \bibnamefont{Bailey}},
  \bibinfo{author}{\bibfnamefont{J.}~\bibnamefont{Balajthy}},
  \bibinfo{author}{\bibfnamefont{P.}~\bibnamefont{Beltrame}},
  \bibinfo{author}{\bibfnamefont{E.~P.} \bibnamefont{Bernard}},
  \bibinfo{author}{\bibfnamefont{A.}~\bibnamefont{Bernstein}},
  \bibinfo{author}{\bibfnamefont{T.~P.} \bibnamefont{Biesiadzinski}},
  \bibinfo{author}{\bibfnamefont{E.~M.} \bibnamefont{Boulton}},
  \bibnamefont{et~al.} (\bibinfo{collaboration}{LUX Collaboration}),
  \href{http://dx.doi.org/10.1103/PhysRevLett.116.161301}{\bibinfo{journal}{Phys.
  Rev. Lett.}, \textbf{\bibinfo{volume}{116}},
  \bibinfo{pages}{161301}\bibinfo{year}{
  (\bibinfo{year}{2016}{\natexlab{a}})}}.

\bibitem[{\citenamefont{Akerib et~al.}(2017{\natexlab{b}})\citenamefont{Akerib,
  Alsum, Aquino, Ara\'ujo, Bai, Bailey, Balajthy, Beltrame, Bernard, Bernstein
  et~al.}}]{LUX2017_Axion}
\bibinfo{author}{\bibfnamefont{D.~S.} \bibnamefont{Akerib}},
  \bibinfo{author}{\bibfnamefont{S.}~\bibnamefont{Alsum}},
  \bibinfo{author}{\bibfnamefont{C.}~\bibnamefont{Aquino}},
  \bibinfo{author}{\bibfnamefont{H.~M.} \bibnamefont{Ara\'ujo}},
  \bibinfo{author}{\bibfnamefont{X.}~\bibnamefont{Bai}},
  \bibinfo{author}{\bibfnamefont{A.~J.} \bibnamefont{Bailey}},
  \bibinfo{author}{\bibfnamefont{J.}~\bibnamefont{Balajthy}},
  \bibinfo{author}{\bibfnamefont{P.}~\bibnamefont{Beltrame}},
  \bibinfo{author}{\bibfnamefont{E.~P.} \bibnamefont{Bernard}},
  \bibinfo{author}{\bibfnamefont{A.}~\bibnamefont{Bernstein}},
  \bibnamefont{et~al.} (\bibinfo{collaboration}{LUX Collaboration}),
  \href{http://dx.doi.org/10.1103/PhysRevLett.118.261301}{\bibinfo{journal}{Phys.
  Rev. Lett.}, \textbf{\bibinfo{volume}{118}},
  \bibinfo{pages}{261301}\bibinfo{year}{
  (\bibinfo{year}{2017}{\natexlab{b}})}}.

\bibitem[{\citenamefont{McKinsey}(2018)}]{McKinsey2018_Ar37}
\bibinfo{author}{\bibfnamefont{D.}~\bibnamefont{McKinsey}},
  \href{http://arxiv.org/abs/arXiv:1803.10110}{\bibinfo{journal}{arXiv}:arXiv:1803.10110\bibinfo{year}{
  (\bibinfo{year}{2018})}}.

\bibitem[{\citenamefont{Akerib et~al.}(2018{\natexlab{a}})\citenamefont{Akerib,
  Alsum, Ara\'ujo, Bai, Bailey, Balajthy, Beltrame, Bernard, Bernstein,
  Biesiadzinski et~al.}}]{LUX2018_Run3PRD}
\bibinfo{author}{\bibfnamefont{D.~S.} \bibnamefont{Akerib}},
  \bibinfo{author}{\bibfnamefont{S.}~\bibnamefont{Alsum}},
  \bibinfo{author}{\bibfnamefont{H.~M.} \bibnamefont{Ara\'ujo}},
  \bibinfo{author}{\bibfnamefont{X.}~\bibnamefont{Bai}},
  \bibinfo{author}{\bibfnamefont{A.~J.} \bibnamefont{Bailey}},
  \bibinfo{author}{\bibfnamefont{J.}~\bibnamefont{Balajthy}},
  \bibinfo{author}{\bibfnamefont{P.}~\bibnamefont{Beltrame}},
  \bibinfo{author}{\bibfnamefont{E.~P.} \bibnamefont{Bernard}},
  \bibinfo{author}{\bibfnamefont{A.}~\bibnamefont{Bernstein}},
  \bibinfo{author}{\bibfnamefont{T.~P.} \bibnamefont{Biesiadzinski}},
  \bibnamefont{et~al.} (\bibinfo{collaboration}{LUX Collaboration}),
  \href{http://dx.doi.org/10.1103/PhysRevD.97.102008}{\bibinfo{journal}{Phys.
  Rev. D}, \textbf{\bibinfo{volume}{97}},
  \bibinfo{pages}{102008}\bibinfo{year}{
  (\bibinfo{year}{2018}{\natexlab{a}})}}.

\bibitem[{\citenamefont{Akerib et~al.}(2017{\natexlab{c}})\citenamefont{Akerib,
  Alsum, Ara\'ujo, Bai, Bailey, Balajthy, Beltrame, Bernard, Bernstein,
  Biesiadzinski et~al.}}]{LUX2017_3DField}
\bibinfo{author}{\bibfnamefont{D.}~\bibnamefont{Akerib}},
  \bibinfo{author}{\bibfnamefont{S.}~\bibnamefont{Alsum}},
  \bibinfo{author}{\bibfnamefont{H.}~\bibnamefont{Ara\'ujo}},
  \bibinfo{author}{\bibfnamefont{X.}~\bibnamefont{Bai}},
  \bibinfo{author}{\bibfnamefont{A.}~\bibnamefont{Bailey}},
  \bibinfo{author}{\bibfnamefont{J.}~\bibnamefont{Balajthy}},
  \bibinfo{author}{\bibfnamefont{P.}~\bibnamefont{Beltrame}},
  \bibinfo{author}{\bibfnamefont{E.}~\bibnamefont{Bernard}},
  \bibinfo{author}{\bibfnamefont{A.}~\bibnamefont{Bernstein}},
  \bibinfo{author}{\bibfnamefont{T.}~\bibnamefont{Biesiadzinski}},
  \bibnamefont{et~al.},
  \href{http://stacks.iop.org/1748-0221/12/i=11/a=P11022}{\bibinfo{journal}{Journal
  of Instrumentation}, \textbf{\bibinfo{volume}{12}},
  \bibinfo{pages}{P11022}\bibinfo{year}{
  (\bibinfo{year}{2017}{\natexlab{c}})}}.

\bibitem[{\citenamefont{Akerib et~al.}(2014)\citenamefont{Akerib, Ara\'ujo,
  Bai, Bailey, Balajthy, Bedikian, Bernard, Bernstein, Bolozdynya, Bradley
  et~al.}}]{LUX2013_FirstResult}
\bibinfo{author}{\bibfnamefont{D.~S.} \bibnamefont{Akerib}},
  \bibinfo{author}{\bibfnamefont{H.~M.} \bibnamefont{Ara\'ujo}},
  \bibinfo{author}{\bibfnamefont{X.}~\bibnamefont{Bai}},
  \bibinfo{author}{\bibfnamefont{A.~J.} \bibnamefont{Bailey}},
  \bibinfo{author}{\bibfnamefont{J.}~\bibnamefont{Balajthy}},
  \bibinfo{author}{\bibfnamefont{S.}~\bibnamefont{Bedikian}},
  \bibinfo{author}{\bibfnamefont{E.}~\bibnamefont{Bernard}},
  \bibinfo{author}{\bibfnamefont{A.}~\bibnamefont{Bernstein}},
  \bibinfo{author}{\bibfnamefont{A.}~\bibnamefont{Bolozdynya}},
  \bibinfo{author}{\bibfnamefont{A.}~\bibnamefont{Bradley}},
  \bibnamefont{et~al.} (\bibinfo{collaboration}{The LUX Collaboration}),
  \href{http://dx.doi.org/10.1103/PhysRevLett.112.091303}{\bibinfo{journal}{Phys.
  Rev. Lett.}, \textbf{\bibinfo{volume}{112}},
  \bibinfo{pages}{091303}\bibinfo{year}{ (\bibinfo{year}{2014})}}.

\bibitem[{\citenamefont{Akerib et~al.}(2018{\natexlab{b}})\citenamefont{Akerib,
  Alsum, Araújo, Bai, Bailey, Balajthy, Beltrame, Bernard, Bernstein,
  Biesiadzinski et~al.}}]{LUX2018_PositionRecon}
\bibinfo{author}{\bibfnamefont{D.}~\bibnamefont{Akerib}},
  \bibinfo{author}{\bibfnamefont{S.}~\bibnamefont{Alsum}},
  \bibinfo{author}{\bibfnamefont{H.}~\bibnamefont{Araújo}},
  \bibinfo{author}{\bibfnamefont{X.}~\bibnamefont{Bai}},
  \bibinfo{author}{\bibfnamefont{A.}~\bibnamefont{Bailey}},
  \bibinfo{author}{\bibfnamefont{J.}~\bibnamefont{Balajthy}},
  \bibinfo{author}{\bibfnamefont{P.}~\bibnamefont{Beltrame}},
  \bibinfo{author}{\bibfnamefont{E.}~\bibnamefont{Bernard}},
  \bibinfo{author}{\bibfnamefont{A.}~\bibnamefont{Bernstein}},
  \bibinfo{author}{\bibfnamefont{T.}~\bibnamefont{Biesiadzinski}},
  \bibnamefont{et~al.},
  \href{http://stacks.iop.org/1748-0221/13/i=02/a=P02001}{\bibinfo{journal}{Journal
  of Instrumentation}, \textbf{\bibinfo{volume}{13}},
  \bibinfo{pages}{P02001}\bibinfo{year}{
  (\bibinfo{year}{2018}{\natexlab{b}})}}.

\bibitem[{\citenamefont{Akerib et~al.}(2017{\natexlab{d}})\citenamefont{Akerib,
  Alsum, Ara\'ujo, Bai, Bailey, Balajthy, Beltrame, Bernard, Bernstein,
  Biesiadzinski et~al.}}]{LUX2017_Kr83m}
\bibinfo{author}{\bibfnamefont{D.~S.} \bibnamefont{Akerib}},
  \bibinfo{author}{\bibfnamefont{S.}~\bibnamefont{Alsum}},
  \bibinfo{author}{\bibfnamefont{H.~M.} \bibnamefont{Ara\'ujo}},
  \bibinfo{author}{\bibfnamefont{X.}~\bibnamefont{Bai}},
  \bibinfo{author}{\bibfnamefont{A.~J.} \bibnamefont{Bailey}},
  \bibinfo{author}{\bibfnamefont{J.}~\bibnamefont{Balajthy}},
  \bibinfo{author}{\bibfnamefont{P.}~\bibnamefont{Beltrame}},
  \bibinfo{author}{\bibfnamefont{E.~P.} \bibnamefont{Bernard}},
  \bibinfo{author}{\bibfnamefont{A.}~\bibnamefont{Bernstein}},
  \bibinfo{author}{\bibfnamefont{T.~P.} \bibnamefont{Biesiadzinski}},
  \bibnamefont{et~al.} (\bibinfo{collaboration}{LUX Collaboration}),
  \href{http://dx.doi.org/10.1103/PhysRevD.96.112009}{\bibinfo{journal}{Phys.
  Rev. D}, \textbf{\bibinfo{volume}{96}},
  \bibinfo{pages}{112009}\bibinfo{year}{
  (\bibinfo{year}{2017}{\natexlab{d}})}}.

\bibitem[{\citenamefont{Akerib et~al.}(2017{\natexlab{e}})\citenamefont{Akerib,
  Alsum, Ara\'ujo, Bai, Bailey, Balajthy, Beltrame, Bernard, Bernstein,
  Biesiadzinski et~al.}}]{LUX2017_Xe127}
\bibinfo{author}{\bibfnamefont{D.~S.} \bibnamefont{Akerib}},
  \bibinfo{author}{\bibfnamefont{S.}~\bibnamefont{Alsum}},
  \bibinfo{author}{\bibfnamefont{H.~M.} \bibnamefont{Ara\'ujo}},
  \bibinfo{author}{\bibfnamefont{X.}~\bibnamefont{Bai}},
  \bibinfo{author}{\bibfnamefont{A.~J.} \bibnamefont{Bailey}},
  \bibinfo{author}{\bibfnamefont{J.}~\bibnamefont{Balajthy}},
  \bibinfo{author}{\bibfnamefont{P.}~\bibnamefont{Beltrame}},
  \bibinfo{author}{\bibfnamefont{E.~P.} \bibnamefont{Bernard}},
  \bibinfo{author}{\bibfnamefont{A.}~\bibnamefont{Bernstein}},
  \bibinfo{author}{\bibfnamefont{T.~P.} \bibnamefont{Biesiadzinski}},
  \bibnamefont{et~al.},
  \href{http://dx.doi.org/10.1103/PhysRevD.96.112011}{\bibinfo{journal}{Phys.
  Rev. D}, \textbf{\bibinfo{volume}{96}},
  \bibinfo{pages}{112011}\bibinfo{year}{
  (\bibinfo{year}{2017}{\natexlab{e}})}}.

\bibitem[{\citenamefont{Dahl}(2009)}]{DahlThesis}
\bibinfo{author}{\bibfnamefont{C.~E.} \bibnamefont{Dahl}}, Ph.D. thesis,
  \bibinfo{school}{Princeton U.} (\bibinfo{year}{2009}),
  \urlprefix\url{https://www.princeton.edu/physics/graduate-program/theses/theses-from-2009/E.Dahlthesis.pdf}.

\bibitem[{\citenamefont{Akerib et~al.}(2016{\natexlab{b}})\citenamefont{Akerib,
  Ara\'ujo, Bai, Bailey, Balajthy, Beltrame, Bernard, Bernstein, Biesiadzinski,
  Boulton et~al.}}]{LUX2016_3H}
\bibinfo{author}{\bibfnamefont{D.~S.} \bibnamefont{Akerib}},
  \bibinfo{author}{\bibfnamefont{H.~M.} \bibnamefont{Ara\'ujo}},
  \bibinfo{author}{\bibfnamefont{X.}~\bibnamefont{Bai}},
  \bibinfo{author}{\bibfnamefont{A.~J.} \bibnamefont{Bailey}},
  \bibinfo{author}{\bibfnamefont{J.}~\bibnamefont{Balajthy}},
  \bibinfo{author}{\bibfnamefont{P.}~\bibnamefont{Beltrame}},
  \bibinfo{author}{\bibfnamefont{E.~P.} \bibnamefont{Bernard}},
  \bibinfo{author}{\bibfnamefont{A.}~\bibnamefont{Bernstein}},
  \bibinfo{author}{\bibfnamefont{T.~P.} \bibnamefont{Biesiadzinski}},
  \bibinfo{author}{\bibfnamefont{E.~M.} \bibnamefont{Boulton}},
  \bibnamefont{et~al.} (\bibinfo{collaboration}{LUX Collaboration}),
  \href{http://dx.doi.org/10.1103/PhysRevD.93.072009}{\bibinfo{journal}{Phys.
  Rev. D}, \textbf{\bibinfo{volume}{93}},
  \bibinfo{pages}{072009}\bibinfo{year}{
  (\bibinfo{year}{2016}{\natexlab{b}})}}.

\bibitem[{\citenamefont{Akerib et~al.}(2017{\natexlab{f}})\citenamefont{Akerib,
  Alsum, Ara\'ujo, Bai, Bailey, Balajthy, Beltrame, Bernard, Bernstein,
  Biesiadzinski et~al.}}]{LUX2017_Yield}
\bibinfo{author}{\bibfnamefont{D.~S.} \bibnamefont{Akerib}},
  \bibinfo{author}{\bibfnamefont{S.}~\bibnamefont{Alsum}},
  \bibinfo{author}{\bibfnamefont{H.~M.} \bibnamefont{Ara\'ujo}},
  \bibinfo{author}{\bibfnamefont{X.}~\bibnamefont{Bai}},
  \bibinfo{author}{\bibfnamefont{A.~J.} \bibnamefont{Bailey}},
  \bibinfo{author}{\bibfnamefont{J.}~\bibnamefont{Balajthy}},
  \bibinfo{author}{\bibfnamefont{P.}~\bibnamefont{Beltrame}},
  \bibinfo{author}{\bibfnamefont{E.~P.} \bibnamefont{Bernard}},
  \bibinfo{author}{\bibfnamefont{A.}~\bibnamefont{Bernstein}},
  \bibinfo{author}{\bibfnamefont{T.~P.} \bibnamefont{Biesiadzinski}},
  \bibnamefont{et~al.} (\bibinfo{collaboration}{LUX Collaboration}),
  \href{http://dx.doi.org/10.1103/PhysRevD.95.012008}{\bibinfo{journal}{Phys.
  Rev. D}, \textbf{\bibinfo{volume}{95}},
  \bibinfo{pages}{012008}\bibinfo{year}{
  (\bibinfo{year}{2017}{\natexlab{f}})}}.

\bibitem[{\citenamefont{Szydagis et~al.}(2013)\citenamefont{Szydagis, Fyhrie,
  Thorngren, and Tripathi}}]{NEST2013}
\bibinfo{author}{\bibfnamefont{M.}~\bibnamefont{Szydagis}},
  \bibinfo{author}{\bibfnamefont{A.}~\bibnamefont{Fyhrie}},
  \bibinfo{author}{\bibfnamefont{D.}~\bibnamefont{Thorngren}},
  \bibnamefont{and} \bibinfo{author}{\bibfnamefont{M.}~\bibnamefont{Tripathi}},
   \href{http://stacks.iop.org/1748-0221/8/i=10/a=C10003}{\bibinfo{journal}{Journal
  of Instrumentation}, \textbf{\bibinfo{volume}{8}},
  \bibinfo{pages}{C10003}\bibinfo{year}{ (\bibinfo{year}{2013})}}.

\bibitem[{\citenamefont{Goetzke et~al.}(2017)\citenamefont{Goetzke, Aprile,
  Anthony, Plante, and Weber}}]{neriX2017_ERCompton}
\bibinfo{author}{\bibfnamefont{L.~W.} \bibnamefont{Goetzke}},
  \bibinfo{author}{\bibfnamefont{E.}~\bibnamefont{Aprile}},
  \bibinfo{author}{\bibfnamefont{M.}~\bibnamefont{Anthony}},
  \bibinfo{author}{\bibfnamefont{G.}~\bibnamefont{Plante}}, \bibnamefont{and}
  \bibinfo{author}{\bibfnamefont{M.}~\bibnamefont{Weber}},
  \href{http://dx.doi.org/10.1103/PhysRevD.96.103007}{\bibinfo{journal}{Phys.
  Rev. D}, \textbf{\bibinfo{volume}{96}},
  \bibinfo{pages}{103007}\bibinfo{year}{ (\bibinfo{year}{2017})}}.

\bibitem[{\citenamefont{\relax
  LUX~Collaboration}(2016)}]{LUX2016_DDCalibration}
\bibinfo{author}{\bibnamefont{\relax LUX~Collaboration}},
  \href{http://arxiv.org/abs/1608.05381}{\bibinfo{journal}{arXiv}:1608.05381\bibinfo{year}{
  (\bibinfo{year}{2016})}}.

\bibitem[{\citenamefont{Akerib et~al.}(2016{\natexlab{c}})\citenamefont{Akerib,
  Ara\'ujo, Bai, Bailey, Balajthy, Beltrame, Bernard, Bernstein, Biesiadzinski,
  Boulton et~al.}}]{LUX2016_Trigger}
\bibinfo{author}{\bibfnamefont{D.}~\bibnamefont{Akerib}},
  \bibinfo{author}{\bibfnamefont{H.}~\bibnamefont{Ara\'ujo}},
  \bibinfo{author}{\bibfnamefont{X.}~\bibnamefont{Bai}},
  \bibinfo{author}{\bibfnamefont{A.}~\bibnamefont{Bailey}},
  \bibinfo{author}{\bibfnamefont{J.}~\bibnamefont{Balajthy}},
  \bibinfo{author}{\bibfnamefont{P.}~\bibnamefont{Beltrame}},
  \bibinfo{author}{\bibfnamefont{E.}~\bibnamefont{Bernard}},
  \bibinfo{author}{\bibfnamefont{A.}~\bibnamefont{Bernstein}},
  \bibinfo{author}{\bibfnamefont{T.}~\bibnamefont{Biesiadzinski}},
  \bibinfo{author}{\bibfnamefont{E.}~\bibnamefont{Boulton}},
  \bibnamefont{et~al.},
  \href{http://dx.doi.org/https://doi.org/10.1016/j.nima.2016.02.017}{\bibinfo{journal}{Nuclear
  Instruments and Methods in Physics Research Section A: Accelerators,
  Spectrometers, Detectors and Associated Equipment},
  \textbf{\bibinfo{volume}{818}}, \bibinfo{pages}{57 }\bibinfo{year}{
  (\bibinfo{year}{2016}{\natexlab{c}})}}, ISSN \bibinfo{issn}{0168-9002}.

\bibitem[{\citenamefont{Aprile et~al.}(2017{\natexlab{b}})\citenamefont{Aprile,
  Aalbers, Agostini, Alfonsi, Amaro, Anthony, Arneodo, Barrow, Baudis,
  Bauermeister et~al.}}]{XENON2017_Modulation}
\bibinfo{author}{\bibfnamefont{E.}~\bibnamefont{Aprile}},
  \bibinfo{author}{\bibfnamefont{J.}~\bibnamefont{Aalbers}},
  \bibinfo{author}{\bibfnamefont{F.}~\bibnamefont{Agostini}},
  \bibinfo{author}{\bibfnamefont{M.}~\bibnamefont{Alfonsi}},
  \bibinfo{author}{\bibfnamefont{F.~D.} \bibnamefont{Amaro}},
  \bibinfo{author}{\bibfnamefont{M.}~\bibnamefont{Anthony}},
  \bibinfo{author}{\bibfnamefont{F.}~\bibnamefont{Arneodo}},
  \bibinfo{author}{\bibfnamefont{P.}~\bibnamefont{Barrow}},
  \bibinfo{author}{\bibfnamefont{L.}~\bibnamefont{Baudis}},
  \bibinfo{author}{\bibfnamefont{B.}~\bibnamefont{Bauermeister}},
  \bibnamefont{et~al.} (\bibinfo{collaboration}{XENON Collaboration}),
  \href{http://dx.doi.org/10.1103/PhysRevLett.118.101101}{\bibinfo{journal}{Phys.
  Rev. Lett.}, \textbf{\bibinfo{volume}{118}},
  \bibinfo{pages}{101101}\bibinfo{year}{
  (\bibinfo{year}{2017}{\natexlab{b}})}}.

\bibitem[{\citenamefont{Abe et~al.}(2018)\citenamefont{Abe, Hiraide, Ichimura,
  Kishimoto, Kobayashi, Kobayashi, Moriyama, Nakahata, Norita, Ogawa
  et~al.}}]{XMASS2018_Modulation}
\bibinfo{author}{\bibfnamefont{K.}~\bibnamefont{Abe}},
  \bibinfo{author}{\bibfnamefont{K.}~\bibnamefont{Hiraide}},
  \bibinfo{author}{\bibfnamefont{K.}~\bibnamefont{Ichimura}},
  \bibinfo{author}{\bibfnamefont{Y.}~\bibnamefont{Kishimoto}},
  \bibinfo{author}{\bibfnamefont{K.}~\bibnamefont{Kobayashi}},
  \bibinfo{author}{\bibfnamefont{M.}~\bibnamefont{Kobayashi}},
  \bibinfo{author}{\bibfnamefont{S.}~\bibnamefont{Moriyama}},
  \bibinfo{author}{\bibfnamefont{M.}~\bibnamefont{Nakahata}},
  \bibinfo{author}{\bibfnamefont{T.}~\bibnamefont{Norita}},
  \bibinfo{author}{\bibfnamefont{H.}~\bibnamefont{Ogawa}}, \bibnamefont{et~al.}
  (\bibinfo{collaboration}{\relax XMASS Collaboration}),
  \href{http://dx.doi.org/10.1103/PhysRevD.97.102006}{\bibinfo{journal}{Phys.
  Rev. D}, \textbf{\bibinfo{volume}{97}},
  \bibinfo{pages}{102006}\bibinfo{year}{ (\bibinfo{year}{2018})}}.

\end{thebibliography}

\end{document}